\documentclass[12pt]{article}

\usepackage[margin=1in]{geometry}
\usepackage{amsmath, amssymb, amsthm, amsfonts}
\usepackage{bm}
\usepackage{graphicx}
\usepackage{natbib}
\usepackage{hyperref}
\usepackage{algorithm, algpseudocode}
\usepackage{xcolor}
\usepackage[normalem]{ulem}


\newcommand{\Y}{\mathbf{Y}}
\newcommand{\Ytilde}{\tilde{\mathbf{Y}}}
\newcommand{\Z}{\mathbf{Z}}
\newcommand{\Hmat}{\mathbf{H}}
\newcommand{\svec}{\mathbf{s}}
\newcommand{\br}{\mathbf{r}}
\newcommand{\y}{\mathbf{y}}

\newcommand{\hvec}{\mathbf{h}}
\newcommand{\thetab}{\boldsymbol{\theta}}
\newcommand{\phib}{\boldsymbol{\phi}}

\newcommand{\bepsilon}{\boldsymbol{\epsilon}}
\newcommand{\W}{\mathbf{W}}
\newcommand{\Gau}{\mathrm{Gau}}
\newcommand{\Pois}{\mathrm{Poisson}}

\newcommand{\E}{\mathbb{E}}
\newcommand{\I}{\mathbf{I}}
\newcommand{\zerov}{\mathbf{0}}

\newcommand{\Yhat}{\hat{\Y}_{0}}

\newcommand{\sphi}{\mathbf{s}_{\phib}}
\newcommand{\rphi}{\mathbf{r}_{\phib}}
\newcommand{\intd}{\mathrm{d}}

\newcommand\independent{\protect\mathpalette{\protect\independenT}{\perp}}
\def\independenT#1#2{\mathrel{\rlap{$#1#2$}\mkern2mu{#1#2}}}

\begin{document}

\title{Spatial Prediction with Guided Diffusion}

\author{
  Andrew Zammit-Mangion\textsuperscript{1}\thanks{Corresponding author: \texttt{a.zammit\_mangion@unsw.edu.au}} \quad
  Lachlan Astfalck\textsuperscript{1} \quad
  Rapha\"el Huser\textsuperscript{2} \\[0.8ex]
  Christopher Nemeth\textsuperscript{3} \quad
  Henry B. Moss\textsuperscript{3} \\[1.4ex]
  \parbox{0.92\textwidth}{\centering\small
    \textsuperscript{1}School of Mathematics and Statistics, UNSW Sydney, Sydney, Australia \\
    \textsuperscript{2}Statistics Program, King Abdullah University of Science and Technology (KAUST), Thuwal, Saudi Arabia \\
    \textsuperscript{3}School of Mathematical Sciences, Lancaster University, Lancaster, United Kingdom
  }
}
\date{}

\maketitle

\begin{abstract}
Spatial prediction is a critically important task in data analyses that involve spatially or spatio-temporally-referenced data. It is commonly carried out using hierarchical models, which comprise a process model describing the latent process of interest, and a data model describing how the process is observed. The target quantity is the predictive distribution, which is the conditional distribution of the latent spatial process given the observed data.  Unfortunately, the predictive distribution is rarely available in closed form and can be difficult to sample from. Here, we show how recent developments in guided diffusion modelling from machine learning can address a wide class of problems in spatial prediction when the likelihood function derived from the data model is known and differentiable, possibly after relaxation. To show the versatility of the approach, we apply it to both Gaussian and non-Gaussian process models and a variety of data models that describe direct partial observations, linear and areal aggregates, counts, right censoring, and area-exceedance functionals. Where possible, we validate the framework against exact and Markov-chain-Monte-Carlo (MCMC)-derived predictive distributions. We demonstrate our approach on several simulation studies and in a controlled experiment of methane-flux inversion over the UK and Ireland. We show that guided diffusion can sample from the correct predictive distribution in a variety of settings, that it can be applied to models where MCMC is intractable, and that it scales well with the dimensionality of the latent process.  

  \vspace{0.3in}
\end{abstract}


\vspace{0.2in}

\section{Introduction}
\label{sec:intro}

A key task in spatial data analysis is to predict a spatial process from various datasets or conditioning information. In studies of biodiversity, for example, a common objective is to infer spatial species dissimilarity from presence/absence records \citep{white2024generative}; in flux inversion studies, the goal is to predict flux (sources and sinks) of a trace gas such as carbon dioxide from both satellite retrievals and in-situ data \citep{zammit2021wombat, bertolacci2024inferring}; in air pollution studies, a common objective is to produce a spatial map of exceedance probabilities \citep{Bolin_2015}; and in ice sheet mass balance studies, one needs to predict the volume and mass of ice from satellite measurements \citep{zammit2015multivariate, martin2016spatial}. These spatial predictions ultimately inform national and international policy and decision making \citep[e.g.,][]{sweet2017global, byrne2023national}.

Spatial prediction can be summarised as follows. Consider a process $\{Y_0(\svec) : \svec \in \mathcal{D}\}$ on some domain $\mathcal{D} \subset \mathbb{R}^d, d \ge 1$ modelled using a \emph{process model} $p(Y_0(\cdot))$, and the availability of information that we can condition on for making inferential statements on $Y_0(\cdot)$. This information is most often in the form of data $\Z$ that are related to the process via a \emph{data model} $p(\Z \mid Y_0(\cdot))$. The target quantity is the predictive distribution given by
\begin{equation}\label{eq:hier-intro}
p(Y_0(\cdot) \mid \Z) \propto p(Y_0(\cdot), \Z) = \underbrace{p(Y_0(\cdot))}_{\text{process model}}\,\underbrace{p(\Z \mid Y_0(\cdot))}_{\text{data model}}.
\end{equation}
If $\Z \mid Y_0(\cdot)$ is linear in $Y_0(\cdot)$ and Gaussian, and $Y_0(\cdot)$ is also Gaussian, then the predictive distribution $p(Y_0(\svec) \mid \Z), \svec \in \mathcal{D},$ is Gaussian, and typically straightforward to sample from. When the process model $Y_0(\cdot)$ is non-Gaussian, direct evaluation of $p(Y_0(\svec) \mid \Z), \svec \in \mathcal{D},$ is not possible, and this predictive distribution is often difficult to sample from. The problem is compounded when the data model is also non-Gaussian and/or non-linear in $Y_0(\cdot)$.

There are several existing approaches for spatial prediction that are tailored to various model choices, each with their own strengths and limitations. A widely applicable approach is Markov chain Monte Carlo (MCMC) which, in many cases, can be used to sample from the target predictive density when this can be evaluated up to a constant of proportionality. Integrated nested Laplace approximations \citep[INLA;][]{Rue2009INLA} is designed for latent Gaussian models. Approximate Bayesian computation \citep[ABC;][]{Sisson2018ABC} precludes the need of pointwise evaluation of process densities or the likelihood function, but scales poorly with the dimension of the summary statistics used in the computation. For specific non-Gaussian processes, bespoke conditional simulators have been developed. In the spatial extremes context, for example, \citet{Dombry2013} develop an MCMC algorithm for conditional simulation of Brown--Resnick max-stable processes. These methods are useful in their specific use cases, but each is tied to a specific process family and data model.

More recently, neural conditional simulation \citep[NCS;][]{Walchessen2024NCS} has been developed to enable fast conditional simulation via score-based diffusion \citep{Song2021SDE}. NCS learns a map from data to the target predictive distribution by training a neural network to directly approximate a score function that is required for conditional simulation. \citet{Walchessen2024NCS} successfully applied their method to the Brown--Resnick spatial process with hundreds of observations, for which MCMC is computationally prohibitive. However, NCS has a practical limitation: the data model must be specified before training the neural network, and cannot be changed at prediction time. If the data model changes or is incorrectly specified, the neural network has to be re-trained. When the data model is unknown until data are observed, or when several data models are likely to be encountered in a given application, this approach is not practical.

The limitation of NCS can be overcome by leveraging \emph{guided diffusion}, where one first constructs a diffusion model to generate unconditionally (rather than conditionally) from $Y_0(\cdot)$, typically using a neural network that estimates a score function. Then, at prediction time, 
one combines the unconditional score function and a likelihood gradient to sample from the predictive distribution. Guided diffusion thus keeps the two factors in \eqref{eq:hier-intro} apart, and only learns the process model in an initial step. Variants of guided diffusion include diffusion posterior sampling (DPS; \citealt{Chung2023DPS}) and $\Pi$GDM \citep{Song2023PiGDM}.  
The flexibility of guided diffusion addresses the misspecification risk inherent in NCS, where the neural network is trained using conditioning data that come from a specific data model or family of data models. Guided diffusion has been explored outside spatial statistics in compressed-sensing imaging, inverse problems, and data assimilation \citep{Kawar2022DDRM, Rozet2023SDA}. Our aim is to adapt and validate its use for the task of spatial prediction.

In this paper we show how guided diffusion can be leveraged as a general approach to making prediction of complex processes from heterogeneous data sources. We demonstrate the utility on both the vanilla Gaussian process and on the amplitude field of the complex-valued nonlinear Ginzburg--Landau equation. For data models, we consider cases where information is in the form of direct partial observations, Poisson counts, right-censored measurements, areal aggregates,  and area-exceedance functionals.
We also demonstrate our framework in a realistic controlled experiment where we infer methane flux from noisy in-situ data. In most cases we compare the predictive distribution to that from MCMC, when it is computationally feasible to do so. We find guided diffusion to be a general and scalable approach to spatial prediction: a single trained process model serves a wide range of data models, and the process model itself need only be available through simulation.

The remainder of the paper is organised as follows. Section~\ref{sec:primer} frames score-based diffusion in the context of spatial prediction, and outlines NCS. Section~\ref{sec:guided} presents guided diffusion for spatial prediction. Section~\ref{sec:experiments} presents results from experiments on Gaussian processes and Ginzburg--Landau fields. Section~\ref{sec:methane} applies the framework in the context of an observing system simulation experiment (OSSE) related to flux inversion. Section~\ref{sec:discussion} concludes.

\section{Score-Based Diffusion for Spatial Prediction}
\label{sec:primer}

Throughout this paper we will consider a discretisation of the process $Y_0(\cdot)$ on a gridding $\mathcal{G}$ of $\mathcal{D}$ comprising $n$ cells, which we denote as $\Y_0 \equiv (Y_0(\svec) : \svec \in \mathcal{G})^\top$. Consider now the forward diffusion process
\begin{equation}
\label{eq:forward-sde}
\intd\Y_t = -\tfrac{1}{2}\beta_t \Y_t \, \intd t + \sqrt{\beta_t}\, \intd\W_t, \qquad t \in [0, T],
\end{equation}
where $\W_t$ is a standard Brownian motion in $\mathbb{R}^n$, $\beta_t > 0$ is a deterministic noise schedule, $t$ denotes diffusion time (which is unrelated to physical time), and where $T > 0$ (we fix $T = 1$ throughout). It is straightforward to show \citep{Song2021SDE} that the conditional distribution of $\Y_t$ given $\Y_0$ is
\begin{equation}
\label{eq:marginal}
\Y_t \mid \Y_0 \sim \Gau(\alpha_t \Y_0,\, \sigma_t^2 \I_n),
\end{equation}
where $\Gau(\boldsymbol{\mu},\boldsymbol{\Sigma})$ is the Gaussian density with mean $\boldsymbol{\mu}$ and covariance matrix $\boldsymbol{\Sigma}$, $\alpha_t = \exp\!\left(-\tfrac{1}{2}\int_0^t \beta_s\, \intd s\right)$ and $\sigma_t^2 = 1 - \alpha_t^2$ (see~\ref{subsec:transition-kernel} for a derivation). At diffusion time $t = T$ the noise schedule is chosen so that $\alpha_T \approx 0$ and $\Y_T \mathrel{\dot\sim} \Gau(\zerov,\I_n)$, regardless of the initial $\Y_0$. Sampling from $p(\Y_0)$ is then equivalent to integrating the corresponding reverse-time stochastic differential equation \citep{Anderson1982} from $\Y_T \sim \Gau(\zerov,\I_n)$ backwards to $\Y_0$:
\begin{equation}
\label{eq:reverse-sde}
\intd\Y_t = \left[-\tfrac{1}{2}\beta_t \Y_t - \beta_t \nabla_{\Y_t} \log p(\Y_t)\right] \intd t + \sqrt{\beta_t}\, \intd\bar{\W}_t,
\end{equation}
where $\bar{\W}_t$ is a Brownian motion in reverse time  (see~\ref{subsec:reverse-diffusion} for a derivation). The only unknown quantity in \eqref{eq:reverse-sde} is the marginal score $\nabla_{\Y_t} \log p(\Y_t)$, which depends on the marginal density of the noised process. 

If $\Y_0$ is Gaussian, then the score $\nabla_{\Y_t} \log p(\Y_t)$ is available in closed form. Specifically, suppose $\Y_0 \sim \Gau(\zerov, \mathbf{K})$ for an $n \times n$ prior covariance matrix $\mathbf{K}$ induced by evaluating a spatial covariance function over the grid $\mathcal{G}$. Marginalising \eqref{eq:marginal} over $\Y_0$ we see that $ \Y_t \sim \Gau\!\left(\zerov,\; \alpha_t^2 \mathbf{K} + \sigma_t^2 \I_n\right),$ which leads to the marginal score
\begin{equation}
\label{eq:gaussian-score}
\nabla_{\Y_t} \log p(\Y_t) = -\left(\alpha_t^2 \mathbf{K} + \sigma_t^2 \I_n\right)^{-1}\Y_t.
\end{equation}
Substituting \eqref{eq:gaussian-score} into the reverse-time SDE \eqref{eq:reverse-sde} yields a linear time-varying system that can be numerically integrated, with no neural network required \citep{moss2026conditioning}. We use this property in the Gaussian-process experiment of Section~\ref{subsec:gp-validation}.

For a general non-Gaussian process, the marginal $p(\Y_t)$, and therefore its score, is intractable, and one must instead estimate it. The usual strategy is to use a neural network for score estimation, which we denote by $\sphi(\Y_t, t) : \mathbb{R}^n \times [0,T] \to \mathbb{R}^n$, where $\phib$ are neural network parameters. The neural network can be trained to be an estimator of the score by minimising the score-matching objective \citep{Vincent2011, Song2021SDE}:
\begin{equation}
\label{eq:dsm}
{\mathcal{L}}_\text{S}(\phib) = \frac{1}{T}\int_{0}^T\E_{\Y_0}\E_{\Y_t \mid \Y_0} \left[\,\left\| \sphi(\Y_t, t) - \nabla_{\Y_t} \log p(\Y_t \mid \Y_0) \right\|^2\,\right]\intd t,
\end{equation}
where $\Y_0 \sim p(\Y_0)$ and $\Y_t \mid \Y_0$ follows from \eqref{eq:marginal}; see~\ref{subsec:Score} for a derivation. The quantity $\nabla_{\Y_t}\log p(\Y_t \mid \Y_0) = -{(\Y_t - \alpha_t \Y_0)}/{\sigma_t^2}$ is numerically unstable for small $\sigma_t$ (that occurs when $t\rightarrow 0$). Therefore, we adopt the approach of training a neural network to estimate the scaled score $-\sigma_t\nabla_{\Y_t}\log p(\Y_t)$ instead \citep{Ho2020DDPM}. Let $\sigma_t\nabla_{\Y_t}\log p(\Y_t \mid \Y_0) \equiv -\br_t$, where the scaled residual $\br_t \equiv  ({\Y_t - \alpha_t \Y_0})/\sigma_t \sim \mathcal{N}(\zerov, \mathbf{I}_n)$. We denote the approximating neural network by $\rphi(\Y_t, t) : \mathbb{R}^n \times [0,T] \to \mathbb{R}^n$, where $\phib$ are neural network parameters. The loss function we minimise is then
\begin{equation}
  \label{eq:dsm2}
  {\mathcal{L}}_\text{R}(\phib) = \frac{1}{T}\int_{0}^T\E_{\Y_0}\E_{\br_t}
    \left[\,\bigl\| \rphi(\alpha_t\Y_0+\sigma_t\br_t,\ t) - \br_t\bigr\|^2\,\right]\intd t.
  \end{equation}
  Once trained, using the same arguments in~\ref{subsec:Score}, $\rphi(\Y_t,t) \approx -\sigma_t\nabla_{\Y_t}\log p(\Y_t)$ for $t\in[0,T]$, and therefore $-\rphi(\Y_t,t) / \sigma_t \approx \nabla_{\Y_t}\log p(\Y_t)$.  In practice \eqref{eq:dsm2} is minimised using Monte Carlo approximation and stochastic gradient descent. Unconditional samples from $p(\Y_0)$ are obtained by integrating \eqref{eq:reverse-sde} backward in time, with $-\rphi(\cdot,\cdot)/\sigma_t$ in place of the true score; see Section~\ref{app:compute} of the Supplementary Material for details. We adopt a neural network to characterise $p(\Y_0)$ in this way in the Ginzburg--Landau experiment of Section~\ref{subsec:gl-prior} and the methane flux inversion experiment of Section~\ref{sec:methane}.

\paragraph{Neural conditional simulation} So far we have discussed how we could use diffusion models to generate \emph{unconditional} samples from $p(\Y_0)$. As we show in Section~\ref{sec:guided}, these samples can be progressively ``guided'' to predictive samples in accordance with different data models. This is in stark contrast to NCS where one instead trains a network to directly generate \emph{conditional} samples (i.e., from $p(\Y_0 \mid \Z)$), through a neural network that targets the conditional score $\nabla_{\Y_t} \log p(\Y_t \mid \Z)$ \citep{Walchessen2024NCS}. NCS is typically implemented using a neural network $\tilde{\mathbf{r}}_{\phib}(\cdot, \cdot, \cdot)$ that also takes $\Z$ as input, and that is trained using the loss
\begin{equation}
    \label{eq:ncs-loss}
    \mathcal{L}_{\mathrm{NCS}}(\phib) = \frac{1}{T}\int_0^T\E_{\Y_0}\E_{\Z \mid \Y_0}\E_{\br_t} \!\left[\,\left\| \tilde{\mathbf{r}}_{\phib}(\alpha_t\Y_0+\sigma_t\br_t, \Z, t) - \br_t \right\|^2\right]\intd t.
    \end{equation}
It is a straightforward extension of the derivation in \ref{subsec:Score} to show that minimising \eqref{eq:ncs-loss}  makes $-\tilde{\mathbf{r}}_{\phib}(\Y_t, \Z, t)/\sigma_t$ target the required conditional score $\nabla_{\Y_t} \log p(\Y_t \mid \Z)$ since ${\Z \independent \Y_t \mid \Y_0}$. For prediction, NCS draws predictive samples by running the reverse SDE with the conditional score in place of the unconditional one. There is one notable drawback of NCS: If, at prediction time, the data do not follow the data model $p(\Z \mid \Y_0)$ used to train the network, then samples from NCS do not target the correct predictive distribution. Retraining for every possible data model can be infeasible in several applications. This is the limitation that the guided diffusion approach, presented in Section~\ref{sec:guided}, removes.

\section{Guided Diffusion for Spatial Prediction with Various Data Models}
\label{sec:guided}

\subsection{Score decomposition and approximate guidance}
\label{subsec:dps}

The reverse-time diffusion \eqref{eq:reverse-sde} samples from whichever distribution the score belongs to. For example, when the score is $\nabla_{\Y_t} \log p(\Y_t)$, reverse diffusion returns draws from the process model $p(\Y_0)$. To sample from the predictive distribution $p(\Y_0 \mid \Z)$ we therefore need to replace the unconditional score in \eqref{eq:reverse-sde} with the conditional score ${\nabla_{\Y_t} \log p(\Y_t \mid \Z)}$. A simple application of Bayes' rule then reveals a decomposition of the score into that of the process model and that of the data model: 
\begin{equation}
\label{eq:score-decomp}
\nabla_{\Y_t} \log p(\Y_t \mid \Z) = \underbrace{\nabla_{\Y_t} \log p(\Y_t)}_{\text{process model score}} + \underbrace{\nabla_{\Y_t} \log p(\Z \mid \Y_t)}_{\text{data model score}}.
\end{equation}
Equation~\eqref{eq:score-decomp} reveals a way forward to defer the choice of data model when training a neural network to estimate the score function. In particular, $\rphi(\cdot,\cdot)$ trained using \eqref{eq:dsm2} to estimate the process model score can be re-used with a wide range of data models at prediction time \citep{lai2025principles}. Importantly, no additional neural network is needed to estimate the data-model score, which in several cases can be directly evaluated.

 While elegant, a drawback of this approach is that, although we have access to $p(\Z \mid \Y_0)$, we do not usually have access to $p(\Z \mid \Y_t)$ since,
\begin{equation}
\label{eq:noised-likelihood}
p(\Z \mid \Y_t) = \int p(\Z \mid \Y_0)\, p(\Y_0 \mid \Y_t)\, \intd\Y_0
\end{equation}
is only available in closed form in rare cases, such as when both $p(\Z \mid \Y_0)$ and $ p(\Y_0 \mid \Y_t)$ are Gaussian. There are several approaches to circumvent this issue, developed largely in the image-generation literature: diffusion posterior sampling \citep[DPS;][]{Chung2023DPS} collapses $p(\Y_0 \mid \Y_t)$ to a point mass at its mean; importance sampling and its sequential Monte Carlo variants approximate the integral directly; and pseudoinverse-guided diffusion \citep[$\Pi$GDM;][]{Song2023PiGDM} replaces $p(\Y_0 \mid \Y_t)$ with a Gaussian surrogate. We describe each in turn.

Tweedie's identity \citep{Efron2011} gives an expression for the conditional expectation of $\Y_0$ given $\Y_t$,
\begin{equation}
\label{eq:tweedie}
\Yhat(\Y_t) \equiv \E[\Y_0 \mid \Y_t] = \frac{1}{\alpha_t}\bigl(\Y_t + \sigma_t^2 \nabla_{\Y_t} \log p(\Y_t)\bigr) \;\approx\; \frac{1}{\alpha_t}\bigl(\Y_t - \sigma_t \rphi(\Y_t, t)\bigr),
\end{equation}
where the approximation uses the estimated score (via the neural network) in place of the true score. The map $\Y_t \mapsto \Yhat(\Y_t)$ can be computed at each diffusion step without any additional forward pass. DPS then approximates \eqref{eq:noised-likelihood} by collapsing the integral \eqref{eq:noised-likelihood} to the Tweedie estimate,
\begin{equation}
\label{eq:dps-approx}
p(\Z \mid \Y_t) \;\approx\; p\!\left(\Z \mid \Yhat(\Y_t)\right),
\end{equation}
neglecting the uncertainty in $\Y_0 \mid \Y_t$, so that the data model score becomes
\begin{equation}
\label{eq:dps-grad}
\nabla_{\Y_t} \log p(\Z \mid \Y_t) \;\approx\; \nabla_{\Y_t} \log p\!\left(\Z \mid \Yhat(\Y_t)\right),
\end{equation}
where the right-hand side is in practice evaluated using automatic differentiation. 

Equation~\eqref{eq:dps-approx} ignores the conditional variance $\mathrm{Cov}(\Y_0 \mid \Y_t)$, which leads DPS to generate conditional samples that are often under-dispersed relative to the true predictive distribution. Alternatively, one could consider instead drawing $\Y_0^{(s)} \sim p(\Y_0 \mid \Y_t)$ and estimating \eqref{eq:noised-likelihood} and its score through importance sampling, but the importance weights tend to be degenerate in even moderately-high dimensional settings. There are sequential-Monte-Carlo variants designed to alleviate this, but this comes at the high computing cost of running a particle filter \citep{Wu2023TDS, Cardoso2024MCGDiff}.

Finally, one could instead adopt $\Pi$GDM, which relaxes the point mass assumption of DPS but that also avoids the degeneracy of importance sampling. Here, one approximates $p(\Y_0 \mid \Y_t)$ using a Gaussian surrogate,
\begin{equation}
\label{eq:pigdm-surrogate}
p(\Y_0 \mid \Y_t) \;\approx\; \Gau\!\left(\Yhat(\Y_t),\, \mathbf{V}_t\right), \qquad \mathbf{V}_t = \left(\mathbf{K}^{-1} + \frac{\alpha_t^2}{\sigma_t^2}\mathbf{I}_n \right)^{-1},
\end{equation}
where recall that $\mathbf{K} \equiv \text{Cov}(\Y_0)$. The expression for $\mathbf{V}_t$ can be derived using the identity  ${\text{Cov}(\Y_0 \mid \Y_t)} = \text{Cov}(\Y_0) - \text{Cov}(\Y_0 , \Y_t)\text{Cov}(\Y_t)^{-1}\text{Cov}(\Y_t , \Y_0)$ and the Sherman--Morrison--Woodbury identity. When $\Y_0$ is Gaussian, this approximation is  exact since, in that case, $\Y_0 \mid \Y_t$ is Gaussian with mean $\Yhat(\Y_t)$ and covariance $\mathbf{V}_t$. In the Gaussian setting of Section~\ref{subsec:gp-validation} we follow \citet{moss2026conditioning} and run the diffusion in whitened coordinates $\Y^{\text{W}}_0 \equiv \mathbf{K}^{-1/2}(\Y_0 - \E[\Y_0])$; applying \eqref{eq:pigdm-surrogate} in those coordinates gives $\mathbf{V}_t = \sigma_t^2\I_n$  (since, as seen in \ref{subsec:transition-kernel}, $\sigma_t^2 + \alpha_t^2 = 1$), which remains exact. When $\text{Cov}(\Y_0)$ is not known, we approximate it with $\varsigma^2\mathbf{I}_n$, where $\varsigma^2$ is the sample variance of the data used for training the neural network. In practice $\varsigma^2 = 1$ since the data are typically standardised before training the network, in which case, once again $\mathbf{V}_t=\sigma_t^2\I_n$.  

Plugging the Gaussian approximation to $p(\Y_0 \mid \Y_t)$ into \eqref{eq:noised-likelihood}, we obtain
\begin{equation}
\label{eq:pigdm-noised}
p(\Z \mid \Y_t) \;\approx\; \int p(\Z \mid \Y_0)\, \Gau\!\left(\Y_0;\, \Yhat(\Y_t),\, \mathbf{V}_t\right) \intd\Y_0,
\end{equation}
where $\Gau(\Y_0; \boldsymbol{\mu},\boldsymbol{\Sigma})$ is the Gaussian density with mean $\boldsymbol{\mu}$ and covariance matrix $\boldsymbol{\Sigma}$ evaluated at $\Y_0$. This quantity, often referred to as the `noised likelihood', is available in closed form with linear-Gaussian data models. In the general non-Gaussian case, a further approximation is needed to evaluate \eqref{eq:pigdm-noised}. We detail the approximations we use when discussing the data models in Sections~\ref{subsec:linear} and \ref{subsec:nonlinear}. Because of the extra approximations required, implementing $\Pi$GDM can be more laborious than implementing DPS. However, as we will see in the experiments of Section~\ref{sec:experiments}, its use generally leads to predictive intervals with empirical coverage closer to the nominal level.

\subsection{Guided diffusion with linear-Gaussian data models}
\label{subsec:linear}

A large fraction of spatial data models used in practice share the linear-Gaussian form
\begin{equation}
\label{eq:linear-obs}
\Z = \Hmat\Y_0 + \bepsilon, \qquad \bepsilon \sim \Gau(\zerov, \sigma_\epsilon^2 \I_m),
\end{equation}
where $\Z \in \mathbb{R}^m$ are noisy data, $\Hmat$ is an $m \times n$ matrix that is (typically) known, $\bepsilon \in \mathbb{R}^m$ is measurement error, and $\sigma^2_\epsilon$ is measurement-error variance. This data model includes the ubiquitous case of direct (partial) observations, where $\Hmat$ is an  incidence matrix. It also includes the important case of linear spatial aggregates, where $\Hmat$ averages the field over disjoint regions: $$(\Hmat\Y_0)_k = |\mathcal{R}_k|^{-1} \sum_{i \in \mathcal{R}_k} Y_0(\svec_i), \quad k = 1,\dots,m,$$ for a partitioning $\{\mathcal{R}_k\}_{k=1}^m$ of the underlying grid into $m$ regions. Another important special case of \eqref{eq:linear-obs} is a soft mass-conservation constraint, where $\Hmat \equiv \hvec^\top = \frac{1}{n} \mathbf{1}^\top$ is a $1 \times n$ incidence vector, and the now scalar observation $Z$ is set to some constraint value with $\sigma^2_\epsilon$ ``small''.  In such cases, the Gaussian log-likelihood evaluated at Tweedie's estimate is, for known $\sigma^2_\epsilon$, $$\log p(\Z \mid \Yhat(\Y_t)) = -\frac{1}{2\sigma^2_\epsilon}(\Z - \Hmat \Yhat(\Y_t))^\top(\Z - \Hmat \Yhat(\Y_t)) + \mathrm{const},$$ and the gradient required for DPS in \eqref{eq:dps-grad} is given by
\begin{equation}
\label{eq:dps-linear-grad}
\nabla_{\Y_t} \log p(\Z \mid \Yhat(\Y_t)) = \frac{1}{\sigma^2_\epsilon}\left(\frac{\partial \Yhat(\Y_t)}{\partial \Y_t}\right)^{\!\top} \Hmat^\top (\Z - \Hmat \Yhat(\Y_t)).
\end{equation}
The Jacobian $\partial \Yhat(\Y_t) / \partial \Y_t = \alpha_t^{-1}\bigl(\I_n - \sigma_t\, \partial \rphi(\Y_t, t) / \partial \Y_t\bigr)$ follows from differentiating \eqref{eq:tweedie}. It involves the derivative of $\rphi(\Y_t, t)$ with respect to $\Y_t$, which in practice is evaluated via automatic differentiation.  For this data model no further approximation is needed when evaluating the noised likelihood \eqref{eq:pigdm-noised} for $\Pi$GDM, since the integral therein is available in closed form.

\subsection{Guided diffusion with non-linear and non-Gaussian data models}
\label{subsec:nonlinear}

It is often the case that spatial data are related to the underlying process in a non-Gaussian or nonlinear way. One of the most common non-Gaussian data models in the spatial setting is a count data model, specifically a conditional Poisson model of the form

\begin{equation}
  \label{eq:poisson-obs}
  \Z \mid \Y_{0} \overset{\mathrm{ind}}{\sim}  \Pois(\rho\Ytilde_0), 
  \end{equation}
  where $\Ytilde_0 \equiv \Hmat g(\Y_0)$, $g : \mathbb{R} \to \mathbb{R}_+$ is a differentiable function applied elementwise, ${\rho > 0}$ is a global intensity parameter, and where the count data $Z_1,\dots,Z_m$ are conditionally independent. The log-Gaussian Cox process is often represented in this way \citep[e.g.,][]{Cseke_2016}, but note that our framework does not require a Gaussian latent field.  As in the Gaussian case, the Jacobian can easily be computed and incorporated into guided samplers.
In Section~\ref{subsec:gp-validation} we let $\Hmat$ be an $m \times n$ incidence matrix identifying which pixels are observed. Details on how we approximate \eqref{eq:pigdm-noised} for $\Pi$GDM with this data model are given in \ref{app:pigdm-poisson}.

Another data model commonly encountered with spatial data is a right-censored observation model. For a known threshold $c$, we observe a noisy value $Z_i = Y_{0,i} + \epsilon_i$ with $\epsilon_i \sim \text{Gau}(0, \sigma^2_\epsilon)$ only when this measurement does not exceed $c$; if $Z_i > c$ the cell is censored in the upper tail and we only know that the measurement lies above $c$. Let $\mathbf{M} \equiv (M_1,\dots, M_n)^\top,$ where $M_i = 1$ corresponds to a censored (i.e., masked) data point. Censoring at $c$ means $\text{Pr}(Z_i > c \mid Y_{0,i}) = \Phi\!\bigl((Y_{0,i} - c)/\sigma_\epsilon\bigr)$, so
\begin{equation}
\label{eq:censor-tail}
\log p(\Z, \mathbf{M} \mid \Y_0)
= \sum_{i: M_i = 0} \left[-\frac{1}{2}\left(\frac{Z_i - Y_{0,i}}{\sigma_\epsilon}\right)^{\!2}\right]
+ \sum_{i: M_i = 1} \log \Phi\!\left(\frac{Y_{0,i} - c}{\sigma_\epsilon}\right),
\end{equation}
up to an additive constant independent of $\Y_0$. This censored log-likelihood is smooth in $Y_{0,i}$ and differentiable. Details on how we approximate \eqref{eq:pigdm-noised} for $\Pi$GDM with this data model are given in \ref{app:pigdm-censoring}.

A related data model we consider is an `area-exceedance' data model, where a single scalar observation $Z = \sum_{i=1}^n \mathbf{1}\{Y_{0,i} > c\}$ counts the number of grid cells in which the field exceeds a known threshold $c$. Such a summary finds use in `what-if' experiments, where one is interested in the distribution of the process conditional on a certain proportion of the domain experiencing extreme events. In this case, we refer to $Z$ as an area-exceedance functional; it is a non-linear scalar functional of the entire field and distinct from the linear spatial aggregates of Section~\ref{subsec:linear}. We can make the likelihood differentiable, as required by guided diffusion, by letting $\sigma_\kappa(Y_{0,i}) \equiv \text{expit}(\kappa(Y_{0,i} - c)), \kappa > 0,$ and endowing $Z$ with a low-variance discrete distribution on $\{0,\ldots,n\}$, centred at $h(\Y_0) \equiv \sum_{i=1}^n \sigma_\kappa(Y_{0,i})$. Here we use a discrete version of the Gaussian distribution:
\begin{equation}
\label{eq:area-exceed-discrete-gauss}
\log \text{Pr}(Z = k \mid \Y_0;\, \kappa, \xi)
= -\frac{(k - h(\Y_0))^2}{2\xi^2}
- \log \sum_{j=0}^n \exp\!\left(-\frac{(j - h(\Y_0))^2}{2\xi^2}\right), \quad k \in \{0,\ldots,n\}.
\end{equation}
The variance $\xi^2$ can be set small; in Section~\ref{subsec:gp-validation} we take $\xi^2 = 0.1$, which concentrates almost all mass within one count of $h(\Y_0)$. Details on how we approximate \eqref{eq:pigdm-noised} for $\Pi$GDM with this data model are given in \ref{app:pigdm-area}.

\section{Experiments}
\label{sec:experiments}

In this section we consider both Gaussian (Section~\ref{subsec:gp-validation}) and non-Gaussian (Section~\ref{subsec:gl-prior}) spatial process models, together with linear-Gaussian and non-Gaussian data models.  On the process side, the Gaussian process model of Section~\ref{subsec:gp-validation} has a score function that is available in closed form, while the non-Gaussian model of Section~\ref{subsec:gl-prior} has an intractable score function that we replace with a trained neural network. On the data side, the linear-Gaussian model \eqref{eq:linear-obs} gives a noised likelihood \eqref{eq:noised-likelihood} that is also available in closed form, while the non-Gaussian data models of Section~\ref{subsec:nonlinear} require the approximations discussed in Section~\ref{subsec:dps}. Details on how the diffusion model was implemented for these experiments and that in Section~\ref{sec:methane} are given in Section~\ref{app:compute} of the Supplementary Material.  In Section~\ref{app:ncs} of the Supplementary Material we also compare guided diffusion to NCS \citep{Walchessen2024NCS}. There, we show that the two approaches are competitive when the data model matches the one used at training time, but that only guided diffusion accommodates a subsequent change in the data model. We use a $32 \times 32$ regular grid ($n = 1024$) to discretise $\mathcal{D} = [0,1]^2$ in the Gaussian process experiment of Section~\ref{subsec:gp-validation}, and $\mathcal{D} = [0,40]^2$ in the Ginzburg--Landau experiment of Section~\ref{subsec:gl-prior}. We let $\Y_0$ be $Y_0(\cdot)$ evaluated on this grid.

\subsection{Gaussian process model}
\label{subsec:gp-validation}

Let $Y_0(\cdot)$ be a mean-zero Gaussian process with a Mat\'ern covariance function of smoothness $\nu = 3/2$, given by
$$
\text{Cov}(Y_0(\svec_i), Y_0(\svec_j)) = \sigma^2_Y \left( 1 + \frac{\sqrt{3}\, \| \svec_i - \svec_j \|}{\ell} \right) \exp\left( -\frac{\sqrt{3}\, \| \svec_i - \svec_j \|}{\ell} \right).
$$
We set the marginal variance $\sigma^2_Y = 1$, and length scale $\ell = 0.2$. Recall from Section~\ref{sec:primer} that, because the gridded $\Y_0$ is Gaussian, the marginal score $\nabla_{\Y_t}\log p(\Y_t)$ is linear in $\Y_t$ and available in closed form. Throughout these Gaussian-process experiments we diffuse the whitened field, as described below \eqref{eq:pigdm-surrogate} and in more detail in \cite{moss2026conditioning}, so that $\mathbf{V}_t = \sigma_t^2\I_n$ in the whitened coordinates and $\sigma_t^2\mathbf{K}$ in the original ones. 

\paragraph{Point-referenced Gaussian observations} We first consider the data model \eqref{eq:linear-obs}.  We simulate $100$ realisations from the underlying spatial process and, from each, we generate a dataset by constructing $\Hmat$ so that a random $10\%$ of the grid cells are observed ($m = 102$) with additive Gaussian measurement-error variance $\sigma_\epsilon^2 = 0.2$. For each realisation we draw $200$ predictive samples in three ways: directly from the closed-form Gaussian predictive distribution \citep[][Chapter 2]{Rasmussen_2006}, and via guided reverse-time diffusion with 500 discretization steps using both DPS and $\Pi$GDM guidance. We use these samples to compute the root-mean-square error (RMSE) between the predictive mean and the true field over all $n = 1024$ grid cells, the empirical coverage of nominal $90\%$ pointwise prediction intervals constructed from equal-tailed empirical quantiles, the field-level energy score (ES), and the averaged per-cell continuous ranked probability score (CRPS) \citep[see][for details]{Gneiting2007}.

\begin{figure}[t!]
\centering
\includegraphics[width=\textwidth]{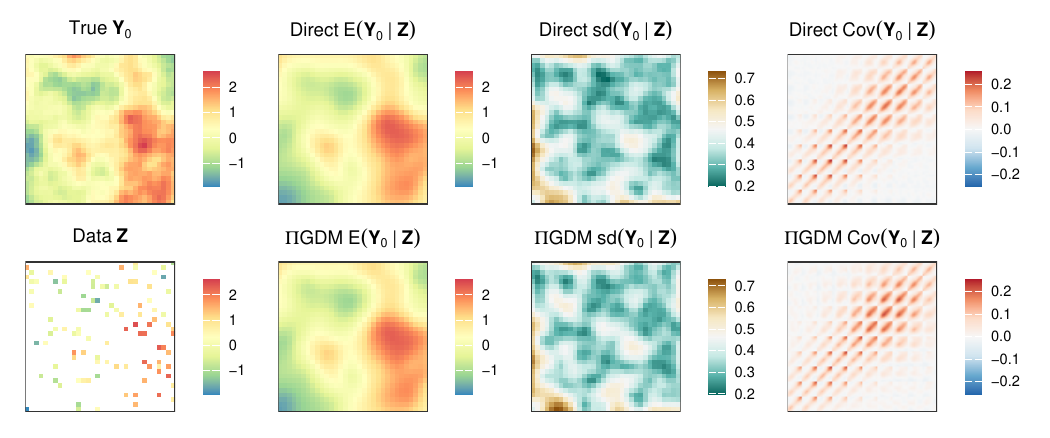}
\caption{Predictive distributions from synthetic Gaussian data generated from a Mat\'ern $3/2$ Gaussian process. Left: true field $\Y_0$ (top) and partial observations $\Z$ (bottom). Centre-left: predictive mean from direct sampling (top) and $\Pi$GDM (bottom). Centre-right: predictive standard deviation from direct sampling (top) and $\Pi$GDM (bottom). Right: predictive covariance of the central $10 \times 10$ block of grid cells from direct sampling (top) and $\Pi$GDM (bottom).}
\label{fig:gp-validation}
\end{figure}

In Figure~\ref{fig:gp-validation} we show results from the first realisation when using $\Pi$GDM. We see that both the marginal and the joint properties of the predictive distribution are accurately captured through guidance. Table~\ref{tab:gp-validation} summarises the diagnostics averaged over the $100$ realisations. There is some loss in fidelity when using DPS over direct sampling from the predictive Gaussian distribution, as expected. On the other hand, guidance using $\Pi$GDM is exact up to time discretisation in this case.

\begin{table}[t!]
  \centering
  \caption{Diagnostics for predictive samples generated directly from the Gaussian predictive distribution (Direct), via DPS guidance, and via $\Pi$GDM guidance, for the Gaussian-process experiment of Section~\ref{subsec:gp-validation}. Entries are means over $100$ realisations, with standard deviation shown in parentheses.}
  \label{tab:gp-validation}
  \vspace{1em}
  
  \begin{tabular}{lcccc}
\hline
Sampler & RMSE & $90\%$ coverage & ES & CRPS \\
\hline
Direct & 0.389\,(0.040) & 0.89\,(0.03) & 8.78\,(0.90) & 0.214\,(0.020) \\
DPS & 0.398\,(0.040) & 0.85\,(0.04) & 9.02\,(0.95) & 0.220\,(0.021) \\
$\Pi$GDM & 0.389\,(0.040) & 0.89\,(0.03) & 8.78\,(0.89) & 0.214\,(0.019) \\
\hline
\end{tabular}

  \end{table}

\begin{figure}[t!]
\centering
\includegraphics[width=\textwidth]{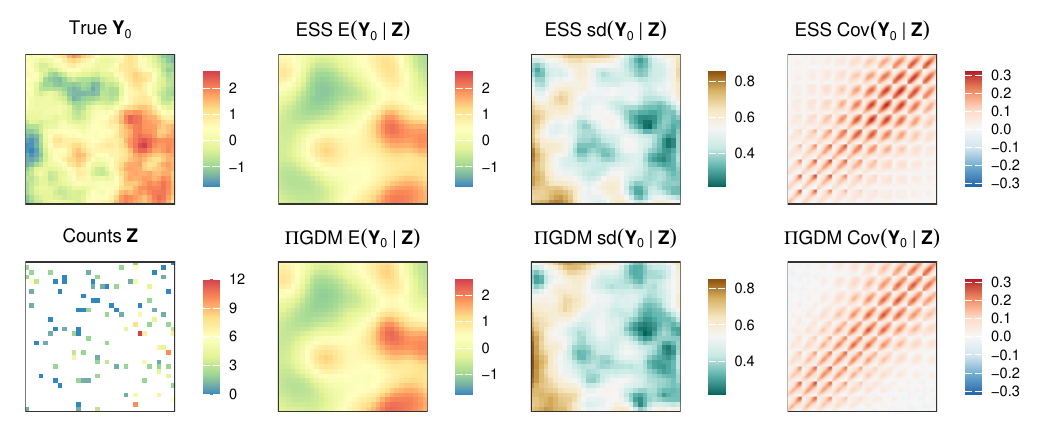}
\caption{Predictive distributions under $10\%$ Poisson-count observations from a Mat\'ern $3/2$ latent Gaussian process. Left: true field $\Y_0$ (top) and observed counts $\Z$ (bottom). Centre-left: predictive mean from ESS (top) and $\Pi$GDM (bottom). Centre-right: predictive standard deviation from ESS (top) and $\Pi$GDM (bottom). Right: predictive covariance of the central $10 \times 10$ block of grid cells from ESS (top) and $\Pi$GDM (bottom).}
\label{fig:gp-poisson}
\end{figure}

  \paragraph{Poisson observations} We now consider the Poisson data model \eqref{eq:poisson-obs} with $g(\cdot)\equiv\exp(\cdot)$, for which the predictive distribution has no closed form. We take the global intensity $\rho = 1$. As in the Gaussian case, we construct $\Hmat$ so that a random $10\%$ of the grid cells are observed, and validate against predictions from an elliptical slice sampler \citep[ESS;][]{Murray2010ESS}. We simulate $100$ realisations from the underlying spatial process and, from each, generate a synthetic count dataset. For each realisation we draw $200$ predictive samples in three ways: via ESS ($200$ independent  chains of $500$ iterations each, retaining the terminal state of each chain as one draw), and via guided reverse-time diffusion with 500 discretization steps using both DPS and $\Pi$GDM guidance. In Figure~\ref{fig:gp-poisson} we show results from the first realisation obtained using $\Pi$GDM, from which we see that the predictive samples recover both the marginal and the joint structure similarly to ESS. Table~\ref{tab:gp-poisson} summarises the diagnostics averaged over the $100$ realisations. The DPS sampler retains the same mild under-dispersion seen in the linear-Gaussian case, while $\Pi$GDM closely matches ESS under these diagnostics.

\begin{table}[t!]
  \centering
  \caption{Diagnostics for predictive samples generated via elliptical slice sampling (ESS), DPS guidance, and $\Pi$GDM guidance, for the Poisson-count experiment of Section~\ref{subsec:gp-validation}. Entries are means over $100$ realisations, with standard deviation shown in parentheses.}  \label{tab:gp-poisson}
  \vspace{1em}

  \begin{tabular}{lcccc}
\hline
Sampler & RMSE & $90\%$ coverage & ES & CRPS \\
\hline
ESS & 0.547\,(0.078) & 0.89\,(0.04) & 12.35\,(1.80) & 0.304\,(0.042) \\
DPS & 0.574\,(0.082) & 0.84\,(0.05) & 13.00\,(1.95) & 0.320\,(0.048) \\
$\Pi$GDM & 0.552\,(0.085) & 0.89\,(0.04) & 12.48\,(1.95) & 0.306\,(0.045) \\
\hline
\end{tabular}

\end{table}

\begin{figure}[t!]
\centering
\includegraphics[width=\textwidth]{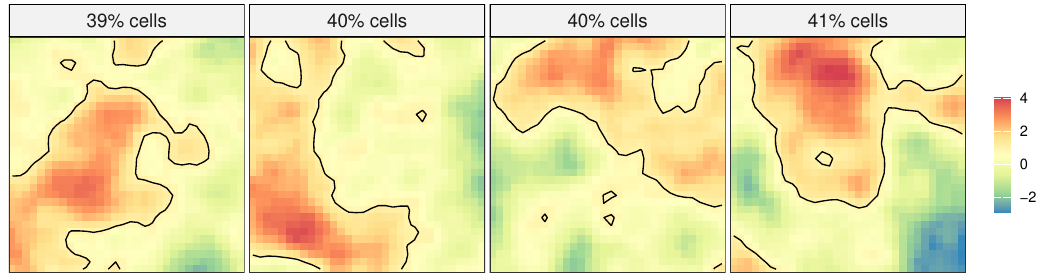}
\caption{Four $\Pi$GDM predictive realisations under the single area-exceedance observation $Z = 410$ cells exceeding $c = 1$. The realisations shown are those at the $2$nd, $34$th, $66$th and $98$th percentiles of the realised exceedance count $\#\{Y_{0,i} > c\}$ across the $200$ predictive draws. Each panel is annotated with the percentage of cells in that realisation that exceed the threshold; the black contour outlines the exceedance region $\{Y_0 > 1\}$. }
\label{fig:gp-area}
\end{figure}

\paragraph{Area-exceedance functional} Here, we consider the area-exceedance data model \eqref{eq:area-exceed-discrete-gauss} with threshold $c = 1$, variance $\xi^2 = 0.1$, sharpness $\kappa = 10$, and a single scalar observation $Z = 410$ (40\% of the $n = 1024$ cells). For each sampler we draw $200$ predictive samples: via guided reverse-time diffusion with 500 discretization steps under DPS and $\Pi$GDM guidance, and via ESS targeting the same discrete-Gaussian likelihood ($200$ independent  chains of $500$ iterations each, retaining the terminal state of each chain as one draw). Table~\ref{tab:gp-area} reports the realised exceedance counts, which are respected by all samplers, although we note that DPS under-disperses relative to ESS on the `hard' count. Figure~\ref{fig:gp-area} shows four $\Pi$GDM realisations from the predictive distribution under the area-exceedance constraint.

When one has a latent Gaussian process model, such as in this section, there are various inferential methods that can be often used to effectively sample from the predictive distribution, such as ESS. The benefit of guided diffusion becomes clearer in the next section, where we apply it to non-Gaussian spatial process models.

\begin{table}[t!]
  \centering
  \caption{Realised exceedance counts under the area-exceedance constraint $Z = 410$ ($c = 1$, $\kappa = 10$) for the experiment in Section~\ref{subsec:gp-validation}. Each sampler draws $200$ independent predictive realisations; entries report mean and standard deviation within parentheses for the exact count $\#(Y_{0,i} > c)$ and for the soft count $h(\Y_0) \equiv \sum_{i=1}^n \sigma_\kappa(Y_{0,i})$ targeted by the discrete-Gaussian surrogate \eqref{eq:area-exceed-discrete-gauss} with $\xi^2 = 0.1$.}
  \label{tab:gp-area}
  \vspace{1em}

  \begin{tabular}{lcc}
\hline
Sampler & $\#(Y_{0,i} > c)$ & $h(\Y_0)$ \\
\hline
ESS & 410.2\,(4.4) & 410.0\,(0.4) \\
DPS & 409.6\,(1.7) & 410.0\,(0.3) \\
$\Pi$GDM & 408.5\,(4.7) & 410.0\,(0.3) \\
\hline
\end{tabular}

\end{table}

\subsection{Ginzburg--Landau process model}
\label{subsec:gl-prior}

In this section we consider a non-Gaussian process model, for which no analytic distribution is available and for which the score function is hence intractable. Specifically, we consider the amplitude field $|A(\svec, \tilde{t})|, \svec \in [0,40] \times [0, 40]$, $\tilde{t} \in [0,20],$ of the two-dimensional complex Ginzburg--Landau equation \citep[CGLE;][]{Aranson2002CGLE},
\begin{align}
\label{eq:cgle}
  \frac{\partial A(\svec, \tilde{t})}{\partial \tilde{t}} = A(\svec, \tilde{t}) + (1 + \iota c_1)\,\nabla^2 A(\svec, \tilde{t}) - (1 - \iota c_3)\,|A(\svec, \tilde{t})|^2 A(\svec, \tilde{t}), \\
  A(s_1+L,s_2,\tilde{t}) = A(s_1,s_2,\tilde{t}), \qquad A(s_1,s_2+L,\tilde{t}) = A(s_1,s_2,\tilde{t}), \nonumber
\end{align}
where $\nabla^2$ is the Laplacian on a torus, $L = 40$ and $\iota \equiv \sqrt{-1}$. The CGLE is a canonical model for pattern formation in oscillatory media, arising in applications as diverse as fluid dynamics and nonlinear optics, with solutions that include spiral waves and travelling fronts. We let $c_1 = c_3 = 1$ on a periodic $64 \times 64$ domain with sides of length $40$. We solve \eqref{eq:cgle}, initialised at $\tilde{t} = 0$ to a small random complex-valued initial field, using the explicit Euler method with $\Delta \tilde{t} = 0.05$ up to physical time $\tilde{t} = 20$. We then compute $|A(\svec, \tilde{t})|$ at $\tilde{t} = 20$, and construct $\Y_0$ by downsampling by a factor of two (so that $\Y_0$ has $n = 1024$ cells). We simulate from \eqref{eq:cgle} in this way, with independent initial conditions, $8{,}000$ times, to yield an ensemble that we split into $7{,}200$ training and $800$ held-out fields for training a neural network. Since this process model has no analytic density, the experiments below cannot be checked against an exact (or MCMC-based) predictive distribution; we therefore rely on predictive diagnostics throughout. Note that $\text{Cov}(\Y_0)$ is not available in closed form; for this reason we use $\mathbf{V}_t = \sigma_t^2 \I_n$ in \eqref{eq:pigdm-surrogate} when approximating the noised likelihood \eqref{eq:pigdm-noised} for $\Pi$GDM.

For $\rphi(\Y_t, t)$ we use a time-conditioned U-Net \citep{Ronneberger2015} with circular padding, trained using the objective \eqref{eq:dsm2}. We consider two data models: an areal linear-Gaussian aggregation model, and an upper-tail censoring model. For 100 different realisations of the ground truth, we generate synthetic observations under each data model, and we draw $200$ predictive samples by discretising reverse diffusion with $500$ steps. For both data models, we report RMSE,  90\% coverage, ES, and CRPS, computed against the true fields and averaged over the $100$ realisations.

\begin{table}[t]
\centering
\caption{Diagnostics from predictive sampling of the Ginzburg--Landau amplitude field under an areal-aggregation and an upper-tail censoring model.  RMSE (of the predictive mean), $90\%$-interval coverage, ES, and CRPS are computed against the true field over all 1024 cells (in the aggregation example) or the censored cells (in the censoring example). Entries are means over 100 realisations, with standard deviation in parentheses.}
\label{tab:gl-experiments}
\vspace{1em}

\begin{tabular}{llcccc}
\hline
Experiment & Sampler & RMSE & $90\%$ coverage & ES & CRPS \\
\hline
Areal aggregates & DPS & 0.091\,(0.023) & 0.855\,(0.037) & 2.05\,(0.53) & 0.045\,(0.012) \\
 & $\Pi$GDM & 0.093\,(0.021) & 0.886\,(0.029) & 2.09\,(0.49) & 0.046\,(0.011)
\\\hline
Censoring ($c=1$) & DPS & 0.010\,(0.002) & 0.828\,(0.049) & 0.10\,(0.01) & 0.005\,(0.001) \\
 & $\Pi$GDM & 0.011\,(0.002) & 0.861\,(0.051) & 0.10\,(0.02) & 0.006\,(0.001)
\\\hline
\end{tabular}
\end{table}

\begin{figure}[t]
\centering
\includegraphics[width=\textwidth]{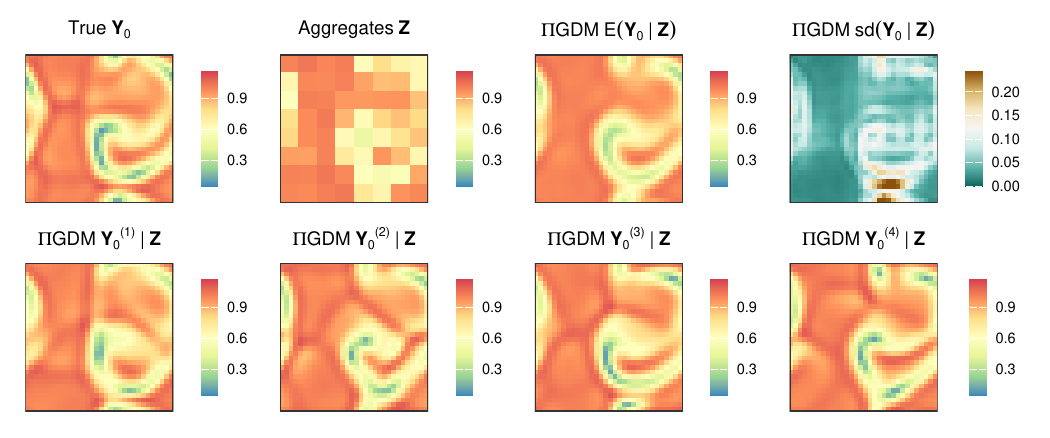}
\caption{Guided $\Pi$GDM predictions of a Ginzburg--Landau amplitude field under $64$ areal aggregates over $4 \times 4$ blocks. Top: true field, the block averages, the predictive mean and the predictive standard deviation obtained using $\Pi$GDM. Bottom: four draws from the predictive distribution using $\Pi$GDM.}
\label{fig:gl-aggregates}
\end{figure}

\paragraph{Aggregated Gaussian observations} In this example we define $\Hmat$ in \eqref{eq:linear-obs} so that $\Z$ is an $8 \times 8$ array of areal aggregates, corrupted with measurement error with standard deviation $\sigma_\epsilon = 0.02$. Figure~\ref{fig:gl-aggregates} shows predictions from $\Pi$GDM guidance for one realisation. Table~\ref{tab:gl-experiments} reports the diagnostics averaged over $100$ independent realisations. There is little difference between DPS and $\Pi$GDM in this example although, as we saw in the Gaussian-process experiments, $\Pi$GDM has better coverage properties. We further assess prediction quality through low-dimensional functionals of each predictive draw relative to the held-out truth. Specifically, we look at the global mean, the global minimum, the global maximum, the proportion of process values below 0.5, and the process in four grid cells (close to $(0, 0)$, $(13, 27)$, $(27, 13)$ and $(39, 39)$).  Figure~\ref{fig:app-pred-functionals-aggregates} of the Supplementary Material summarises the differences between the true underlying functionals and the sampled functionals for 100 different realisations, with DPS and $\Pi$GDM under areal aggregation. The differences are centred near zero with a small spread across all functionals for both samplers, indicating good recovery of these functionals with a slight bias in the predicted global mean.

\begin{figure}[t]
\centering
\includegraphics[width=\textwidth]{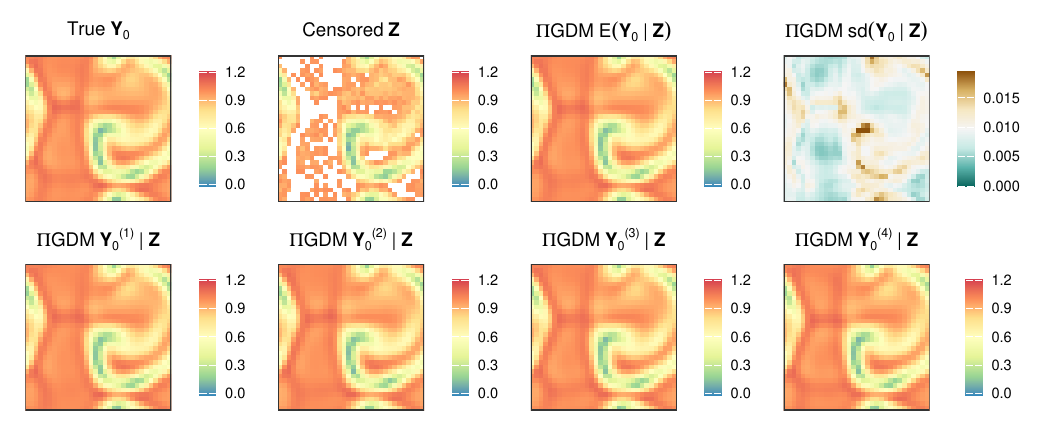}
\caption{Guided $\Pi$GDM predictions of a Ginzburg--Landau amplitude field under upper-tail censoring at $c = 1$. Top: true field, the uncensored observations (with censored cells shown in white), the predictive mean and the predictive standard deviation. Bottom: four draws from the predictive distribution using $\Pi$GDM.}
\label{fig:gl-censored}
\end{figure}

\paragraph{Censored data} In this example we assume that the  measurements are censored in their upper tail at $c = 1$ and that the uncensored cells are observed with measurement error having standard deviation $\sigma_\epsilon = 0.03$. Note that, because of measurement error, a cell may be censored even when its underlying process value lies below $c$, and vice versa. Figure~\ref{fig:gl-censored} shows predictions from $\Pi$GDM guidance for one realisation. Table~\ref{tab:gl-experiments} reports the diagnostics averaged over $100$ independent realisations with, again, $\Pi$GDM achieving closer-to-nominal coverage than DPS. Figure~\ref{fig:app-pred-functionals-censored} of the Supplementary Material shows the differences of the functionals analysed in the areal-observation case. The predictions again appear to be consistent with the underlying truth, with the global mean for $\Pi$GDM showing a small positive bias while the remaining functional differences stay close to zero. 

\vspace{0.1in}

Note that both experiments in this section used the same neural network for estimating the process model score function; only the guidance gradient differed between the two. With NCS, a separate network would need to be trained for each data model. Section~\ref{app:ncs} of the Supplementary Material illustrates potential problems that could arise if NCS is used in a setting where the data model is misspecified.

\section{Observing system simulation experiment of methane-flux inversion over the UK and Ireland}
\label{sec:methane}

We now explore the potential of guided diffusion in a realistic controlled setting, by applying it to atmospheric methane-flux inversion over the UK and Ireland. The particular inversion problem we consider can be tackled using MCMC which, however, scales poorly with spatial dimension. We show that, here, guided diffusion can be used to obtain predictive inferences that are similar to those from MCMC, while offering the possibility to scale up to higher spatial resolutions.

\subsection{Trace-gas inversion using guided diffusion}
\label{subsec:methane-setup}

 The specific atmospheric trace-gas inversion problem we consider is to infer a spatial flux field $\mathbf{Y}_0 \equiv (Y_0(\svec) : \svec \in \mathcal{G})^\top$ on a gridded domain $\mathcal{G}$ from observations of the atmospheric mole fraction recorded at a small number of ground stations. For a long-lived gas such as methane, the mole fraction is, to first order, a known linear functional of the flux: meteorology and transport map the flux at every grid cell to the mole fraction at each station and time point through a sensitivity (``footprint'') operator computed by a Lagrangian particle-dispersion model \citep[here we use the UK Met Office Numerical Atmospheric-dispersion Modelling Environment,  NAME;][]{Jones2007}. The data of methane mole fraction  $\mathbf{Z} \in \mathbb{R}^m$, which are recorded at stations around the UK at fine time intervals, are assumed to follow the data model
\begin{equation}
\label{eq:methane-obs}
\mathbf{Z} \mid \Y_0 \sim \Gau\!\left(\Hmat\Y_0,\; \boldsymbol{\Sigma}_{{\epsilon}}\right), \qquad \boldsymbol{\Sigma}_{{\epsilon}} = \boldsymbol{\Sigma}_\zeta + \boldsymbol{\Sigma}_{\text{o}},
\end{equation}
where $\mathbf{H} \in \mathbb{R}^{m \times n}$ is the (NAME-derived) transport operator mapping the $n$ flux cells to the $m$ station--time observations, $\boldsymbol{\Sigma}_\zeta \in \mathbb{R}^{m \times m}$ is a covariance matrix accounting for transport-model error, and $\boldsymbol{\Sigma}_{\text{o}}$ is a diagonal measurement-error covariance matrix. For illustration purposes, here we assume that the transport-model error covariance is known from previous studies. Following \citet{ZammitMangion2016}, we take $\boldsymbol{\Sigma}_\zeta$ to be a space-time separable covariance matrix, specified through $\boldsymbol{\Sigma}_\zeta^{-1} = \sigma_\zeta^{-2}\,\bigl(\mathbf{Q}^t_a \otimes (\mathbf{R}^s_d)^{-1}\bigr)$, where $\mathbf{Q}^t_a$ is the precision of a first-order autoregressive process in time with lag-one correlation $a = 0.9$ on the two-hourly observation grid, $\mathbf{R}^s_d$ is an exponential spatial correlation matrix across the stations with range $d = 2.5^\circ$, and $\sigma_\zeta = 10$~ppb is a scale parameter. The measurement-error covariance $\boldsymbol{\Sigma}_{\text{o}}$ is diagonal with diagonal elements $1$~ppb$^2$. Model \eqref{eq:methane-obs} is the linear-Gaussian model of Section~\ref{subsec:linear}, but with a non-diagonal covariance matrix $\text{Cov}(\boldsymbol{\epsilon}) = \boldsymbol{\Sigma}_{{\epsilon}}$.

\begin{figure}[t]
\centering
\includegraphics[width=\textwidth]{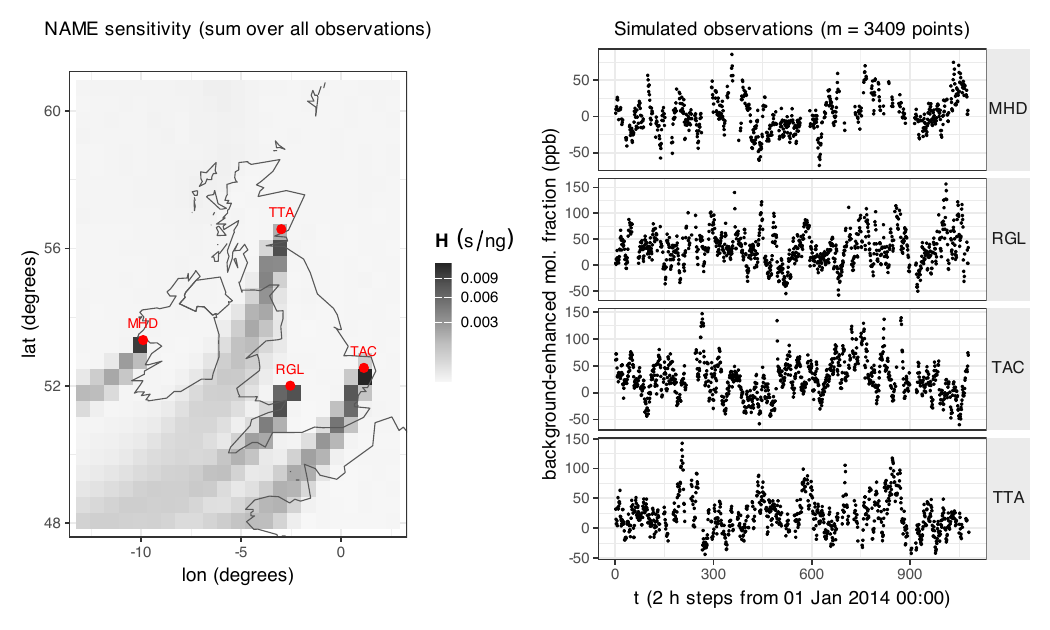}
\caption{Set-up for the methane controlled experiment. Left: the UK and Ireland flux domain with the four station locations (MHD, RGL, TAC, TTA) and the NAME sensitivity plumes comprising $\Hmat$ at the first time point (corresponding to  January 01 2014 at 00:00). Right: the four simulated `above background' methane mole-fraction time series ($m = 3409$ observations, with a sampling interval of 2 h) used as $\Z$ in Section~\ref{sec:methane}.}
\label{fig:methane-setup}
\end{figure}

The first setup we consider is the same as that in \citet{ZammitMangion2016}. The gridded flux domain $\mathcal{G}$ is a $0.7^\circ \times 0.5^\circ$ longitude--latitude grid over the UK and Ireland with $n = 122$ land cells (arising from downsampling the native transport-model resolution by a factor of two). The true (but assumed unknown) flux field is taken to be the UK and Ireland methane inventory (a combination of the NAEI and EDGAR inventories; \citealp{Ganesan2015}), whose domain total at this resolution is $2.73$~Tg\,yr$^{-1}$ ($86{,}590$~g\,s$^{-1}$). We use the transport model NAME to construct the matrix $\mathbf{H}$, so that the $(i,j)$th element of $\mathbf{H}$ corresponds to the sensitivity of the $i$th observation to the flux in the $j$th grid cell \citep[sensitivities are computed by running NAME backwards in time for each space-time observation; see ][for details]{Ganesan2015}. We then forward-simulate mole-fraction observations through \eqref{eq:methane-obs} at four stations: Mace Head, Ireland (MHD); Ridge Hill, England (RGL); Tacolneston, England (TAC); and Angus, Scotland (TTA). Simulations are done at two-hourly resolution over January--March 2014 using the true meteorological conditions during that period. A missingness pattern corresponding to that in actual data is applied, giving a total of $m = 3409$ space-time observations. The domain, station footprints, and simulated time series are shown in Figure~\ref{fig:methane-setup} (figure adapted from \citealt{ZammitMangion2016}).

Following \citet{ZammitMangion2016}, we model the flux field as a Box--Cox spatial process \citep{deOliveira1997}. Let
\begin{equation}
\label{eq:boxcox}
g_\lambda(y) \equiv \begin{cases} (y^\lambda - 1)/\lambda, & \lambda \neq 0, \\ \log y, & \lambda = 0,\end{cases}
\end{equation}
denote the Box--Cox transformation. We model $g_\lambda(\Y_0)$ as a truncated Gaussian vector with density
\begin{equation}
\label{eq:bc-density}
p(g_\lambda(\Y_0) \mid \mathbf{b}, \tau, \thetab, \lambda) \propto \exp\!\left(-\frac{\tau}{2}\left(g_\lambda(\Y_0) - \mathbf{X}\mathbf{b}\right)^\top \mathbf{R}_{\thetab}^{-1} \left(g_\lambda(\Y_0) - \mathbf{X}\mathbf{b}\right)\right) T_\lambda(g_\lambda(\Y_0)),
\end{equation}
and express the density of $\Y_0$ as $\Y_0 \sim \text{BC}\!\left(\mathbf{X}\mathbf{b},\, \tau^{-1}\mathbf{R}_{\thetab},\, \lambda\right)$. Here, $T_\lambda(\tilde{y}) \equiv 1(\tilde{y} > -1/\lambda)$ for $\lambda > 0$, and $1(\tilde{y} < -1/\lambda)$ for $\lambda < 0$ and, like $g_\lambda(\cdot)$, is applied elementwise to its vector argument. The matrix $\mathbf{X} \in \mathbb{R}^{n \times p}$ contains covariates, $\mathbf{b} \in \mathbb{R}^{p}$ is the corresponding vector of regression coefficients, $\tau$ is a marginal precision parameter, and $\mathbf{R}_{\thetab} \equiv (\exp(-\theta_1\|\svec_i - \svec_j\|^{\theta_2}) : i,j = 1,\dots,n)$ is a powered-exponential correlation matrix with $\theta_1 > 0$ and $0 < \theta_2 < 2$. As covariates we use two indicator functions that split the domain at $56.4^\circ$N, to allow for the much lower emissions typical of the Scottish Highlands; the two elements of $\mathbf{b}$ are therefore the mean levels of $g_\lambda(\Y_0)$ outside and inside the Highlands, respectively.

We equip the unknown parameters in the flux field with prior distributions. Since $\mathbf{b}$ and $\tau$ are defined on the Box--Cox scale, which changes with $\lambda$, these parameters cannot be assigned independent prior distributions. We therefore instead assign prior distributions to these parameters on their natural scale. Specifically, we define a prior distribution for the transformed parameters:
\begin{equation}
  \label{eq:methane-reparam}
\tilde{b}_1 = g_\lambda^{-1}(b_1), \qquad \tilde{b}_2 = g_\lambda^{-1}(b_2), \qquad c_\text{V} = \tau^{-1/2} (\tilde{b}_1^\lambda)^{-1}
\end{equation} 
where the expression for the coefficient of variation $c_\text{V}$ follows from $$\mathrm{sd}(g_\lambda(Y_0(\svec))) \approx {\tilde{b}_1^{\lambda - 1}\mathrm{sd}(Y_0(\svec)) = \tilde{b}_1^\lambda c_\text{V}}.$$ We elicit the prior distribution by first running MCMC on the EDGAR/NAEI inventory treated as a realisation of $\Y_0$, under uninformative priors for the parameters (MCMC is computationally tractable when $n=122$). We use the posterior distribution of the parameters from that run to construct a prior distribution for the flux inversion from mole fraction data. Figure~\ref{fig:methane-prior-hyper} of the Supplementary Material shows these elicited parameter prior densities. The full hierarchical model we use in the inversion is given by
\begin{align}
\label{eq:methane-hier}
\textrm{Data model} & \qquad (\Z \mid \Y_0) \sim \Gau\!\left(\Hmat\Y_0,\, \boldsymbol{\Sigma}_\epsilon\right), \nonumber \\
\textrm{Process model} & \qquad (\Y_0 \mid \mathbf{b}, \tau, \thetab, \lambda) \sim \text{BC}\!\left(\mathbf{X}\mathbf{b},\, \tau^{-1}\mathbf{R}_{\thetab},\, \lambda\right), \nonumber \\
\textrm{Parameter model} & \qquad \lambda \sim \Gau(0.168, 0.068^2)\,I(0 < \lambda < 0.5), \nonumber \\
 & \qquad \log \theta_1 \sim \Gau(-0.005, 0.387^2), \quad~ \mathrm{logit}(\theta_2/2) \sim \Gau(0.290, 0.456^2), \nonumber \\
 & \qquad \log \tilde{b}_1 \sim \Gau(6.334, 0.260^2), \qquad \log \tilde{b}_2 \sim \Gau(4.540, 0.464^2), \nonumber \\
 & \qquad \log c_V \sim \Gau(-0.366, 0.168^2),
\end{align}
with $\mathbf{b}$ and $\tau$ obtained from $(\lambda, \tilde{b}_1, \tilde{b}_2, c_{\text{V}})$ through \eqref{eq:methane-reparam}. The logarithmic and logit transformations respect the parameter constraints: $\tau, \theta_1 > 0$ and $\theta_2 \in (0,2)$.

Marginalising \eqref{eq:methane-hier} over $(\mathbf{b}, \tau, \thetab, \lambda)$ yields a non-Gaussian process model $p(\Y_0)$ that is available only implicitly through simulation: We draw parameters from the parameter prior distribution, and then draw $\Y_0$ from the corresponding Box--Cox process in order to provide a training data set of size $3.6\times 10^4$ for training the score network. The mole-fraction data model \eqref{eq:methane-obs} is not used for training, it only enters through the guidance gradient when making inference. Figure~\ref{fig:methane-prior} of the Supplementary Material  shows unconditional draws from the trained generative model.

\subsection{Guided predictions and comparison with MCMC}
\label{subsec:methane-results}

\begin{figure}[t!]
\centering
\includegraphics[width=\textwidth]{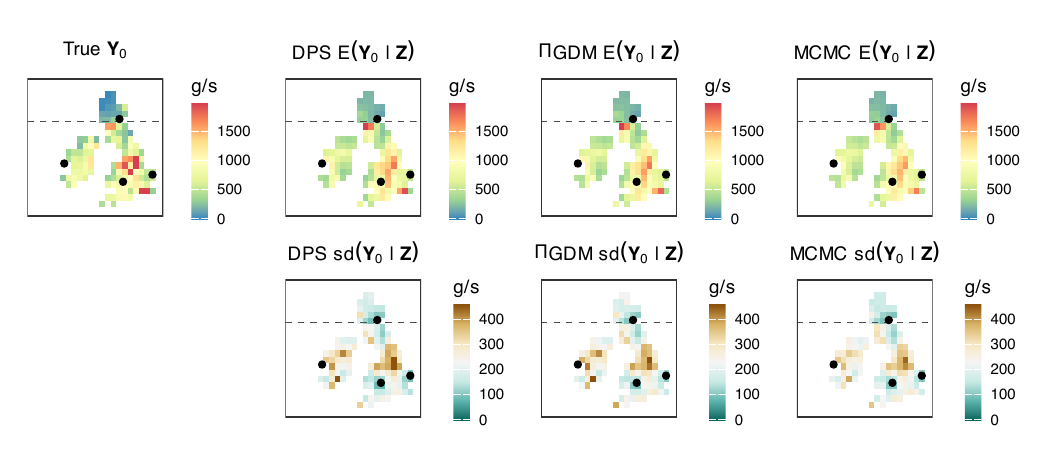}
\caption{Predictive mean and predictive standard deviation of the methane flux field for DPS, $\Pi$GDM, and MCMC (g\,s$^{-1}$, $n=122$). The leftmost panel shows the methane flux inventory map, here treated as the underlying truth. Station locations are marked with dots, while  the dashed line is the $56.4^\circ$N covariate split.} 
\label{fig:methane-means-and-se}
\end{figure}

\begin{figure}[t!]
\centering
\includegraphics[width=\textwidth]{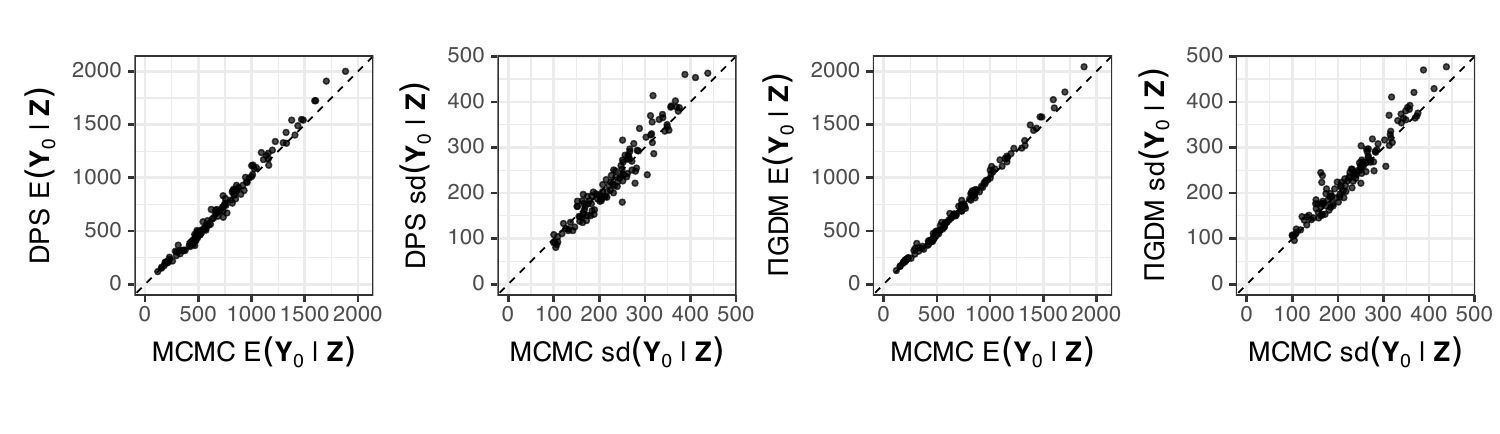}
\caption{Per-cell agreement between each guided sampler and MCMC over the $122$ cells. From left to right: DPS predictive means, DPS predictive standard deviations, $\Pi$GDM predictive means, and $\Pi$GDM predictive standard deviations. Dashed lines indicate the $1\!:\!1$ reference.}
\label{fig:methane-scatter}
\end{figure}

We draw a sample of size $500$ from the predictive distribution using guided diffusion with $500$ discretization steps (setting $\mathbf{V}_t = \sigma_t^2 \I_n$ after data standardisation for $\Pi$GDM), and compare it to the Hamiltonian-Monte-Carlo-within-Gibbs sampling of $\Y_0$ and the model parameters in \eqref{eq:methane-hier} using the code of \citet{ZammitMangion2016}. Table~\ref{tab:methane-diag} shows that both guided predictions approximately match MCMC on several diagnostics computed against the truth: RMSE, per-grid-cell correlation, CRPS, and ES. The $90\%$-interval coverages are similar across the three samplers, and all three domain-total estimates are biased slightly upward, with the MCMC total closest to the truth. The predictive mean and standard deviation maps in Figure~\ref{fig:methane-means-and-se} are visually almost indistinguishable across the three samplers, and the guided and MCMC predictions agree more closely with each other than either agrees with the truth; see Figure~\ref{fig:methane-scatter}.

\begin{table}[t!]
\centering
\caption{Methane flux-field diagnostics for the two guided samplers (DPS and $\Pi$GDM) and MCMC targeting the same hierarchical model \eqref{eq:methane-hier} (Section~\ref{subsec:methane-results}). RMSE, correlation, $90\%$-interval coverage, CRPS, and ES are computed against the true flux field over the $n = 122$ cells; the final column gives the domain-total predictive mean and predictive standard deviation. The true domain total is $86{,}590$~g\,s$^{-1}$ ($2.73$~Tg\,yr$^{-1}$); coverage target is $0.90$. The diagnostics are computed from a sample of size $500$.} \vspace{0.1in}
\label{tab:methane-diag}
\begin{tabular}{lcccccc}
\hline
Method & RMSE & corr & $90\%$ cov.\ & CRPS & ES & total $\pm$ sd (g\,s$^{-1}$) \\
       & (g\,s$^{-1}$) & (truth) & & & & \\
\hline
DPS (reverse SDE) & $298$ & $0.80$ & $0.80$ & $161$ & $2{,}336$ & $87{,}318 \pm 2{,}547$ \\
$\Pi$GDM (reverse SDE) & $296$ & $0.80$ & $0.82$ & $159$ & $2{,}318$ & $87{,}367 \pm 2{,}938$ \\
MCMC & $297$ & $0.80$ & $0.80$ & $158$ & $2{,}336$ & $86{,}644 \pm 3{,}015$ \\
\hline
\end{tabular}

\end{table}

\begin{figure}[t!]
  \centering
  \includegraphics[width=\textwidth]{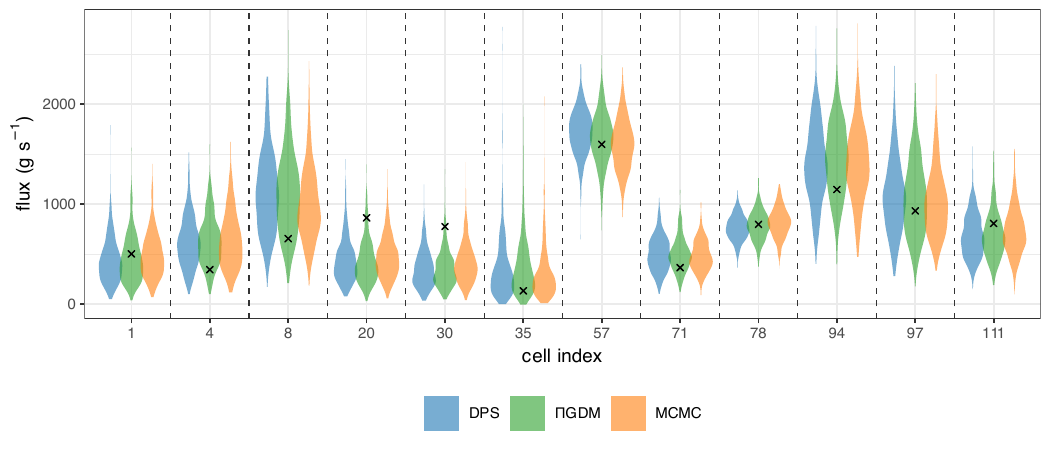}
  \caption{Marginal predictive distributions of the flux (g\,s$^{-1}$) at twelve randomly chosen grid cells, comparing the guided samplers (blue and green) with MCMC (orange). The black cross marks the true (inventory) flux.}
  \label{fig:methane-violin}
  \end{figure}

Figure~\ref{fig:methane-violin} compares the marginal predictions of the guided and MCMC samplers at twelve randomly selected cells. The predictive distributions are closely matched in location, spread, and skewness at every cell.  Table~\ref{tab:methane-countries} of the Supplementary Material reports region-level totals. The guided (DPS and $\Pi$GDM) and MCMC predictive distributions agree closely with one another: all three reproduce the English and Welsh totals, over-estimate the Scottish total, and slightly under-estimate the Northern Irish and Irish totals. These territorial biases sum to the domain-total offset, indicating a coherent spatial redistribution of flux, as is typically seen in studies of this type.

\begin{figure}[t]
\centering
\includegraphics[width=\textwidth]{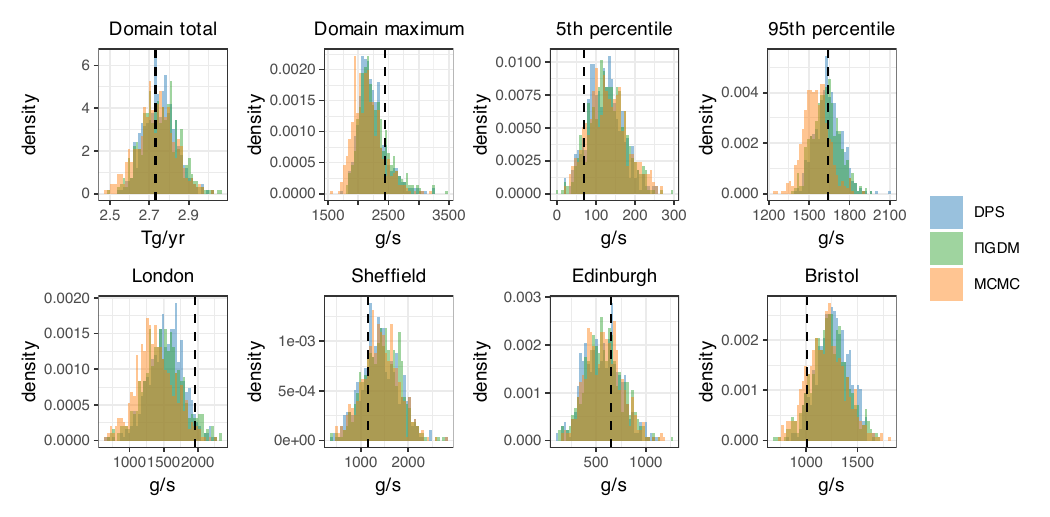}
\caption{Predictive distributions of eight low-dimensional functionals of the methane flux field ($n = 122$), for DPS (blue), $\Pi$GDM (green), and MCMC (orange, labelled MCMC); the dashed line marks the true (inventory) value. Top row: domain total, domain maximum, and the within-draw spatial $5$th and $95$th percentiles. Bottom row: marginal flux at grid cells nearest to London, Sheffield, Edinburgh, and Bristol.}
\label{fig:methane-functionals}
\end{figure}

We assess prediction quality through low-dimensional functionals of the predictive distribution. Figure~\ref{fig:methane-functionals} compares, for DPS, $\Pi$GDM, and MCMC, the domain total and maximum, the spatial $5$th and $95$th percentiles, and the marginal flux at four cities. The guided and MCMC distributions corroborate on all quantities except the upper percentile where the approximation done with guided diffusion is likely leading to slight over-estimation of the fluxes.

\subsection{Computational efficiency}\label{sec:comp_efficiency}

Implementing MCMC is possible in this application where the data model is Gaussian, and where the hierarchical model can be sampled from using Gibbs steps. The utility of guidance here lies in speed and scalability: the score needed in the reverse diffusion process is estimated using feedforward operations and has cost that scales linearly with the dimension of $\Y_0$. On the other hand, MCMC and other Bayesian algorithms applied to spatial processes of this kind require factorising a matrix of size $n \times n$ several times. To assess scalability, we consider the seconds per effective sample of the total methane flux for both approaches under different spatial resolutions: $n = 122$ (the downsampled case considered above), $n = 487$ (the native resolution), and  $n = 1948$, where the flux map and the sensitivity plumes at this higher resolution are obtained through piecewise linear interpolation. Each resolution has its own simulated mole-fraction data $\Z$, generated from the flux field at that resolution under the same transport-error model \eqref{eq:methane-obs}; the number of observations $m = 3409$ is held fixed across the three cases. At the higher resolutions we use the same model in \eqref{eq:methane-hier} but with $\tilde{b}_1$ and $\tilde{b}_2$ appropriately rescaled since they describe median fluxes in a single grid cell, and hence need to be adapted to the cell area. For guided diffusion, every sample is an effective sample (albeit from an approximate posterior distribution) while, for MCMC, chain correlation is taken into account when calculating effective sample size. We run guided diffusion  on an NVIDIA H100 GPU, and MCMC with ten chains in parallel on ten Xeon(R) Gold 6330 CPU cores. 

\begin{figure}[t!]
\centering
\includegraphics[width=0.85\textwidth]{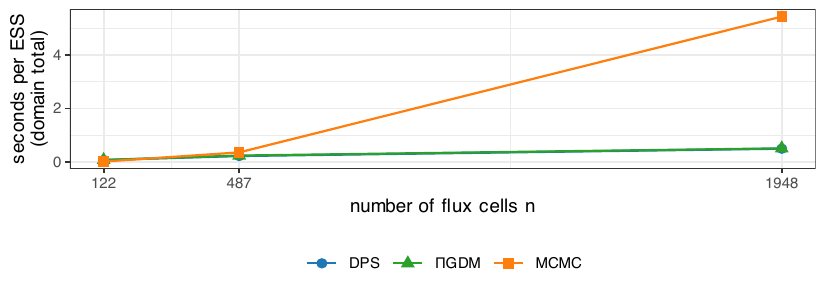}
\caption{Computational scaling of guided diffusion and MCMC for the methane inversion shown as the time required per effective sample of domain total against the number of flux cells $n$, where $n = 122$, $487$ and $1948$. Guided-diffusion timings are from running reverse diffusion on a GPU, while MCMC timings are from running ten parallel chains on ten CPU cores.}
\label{fig:methane-scaling}
\end{figure}

$\Pi$GDM required $0.071$, $0.234$ and $0.506$ seconds per effective sample of the domain total flux for $n = 122$, $487$ and $1948$, respectively. MCMC required $0.010$, $0.353$ and $5.45$ seconds, respectively. Therefore, MCMC is more computationally efficient in low dimension, but its cost grows dramatically faster than that of guided prediction because of the matrix factorisation required at each iteration; it is roughly eleven times more expensive per effective sample at $n = 1948$, and it is not practical at higher resolutions. Figure~\ref{fig:methane-scaling} shows the time required per effective sample of the domain total flux against $n$ for both methods, and clearly shows the potential of guided diffusion in high-resolution flux inversion. Functional diagnostics and predictions at these higher resolutions are shown in Figures \ref{fig:hires2-functionals}--\ref{fig:hires2-mean-sd} and \ref{fig:ultra-functionals}--\ref{fig:ultra-mean-sd} of the Supplementary Material.

\section{Discussion}
\label{sec:discussion}

In this paper we apply guided diffusion to spatial prediction, and show its applicability to a wide variety of process and data models that include right censoring and area-exceedance functionals. Where feasible, we show that guided diffusion recovers predictive distributions that are close to what one can obtain from MCMC. For the field of spatial statistics, the implications are threefold. First, the method is very general, and can be applied whenever one is able to simulate from the process model and when one can construct a differentiable likelihood function from a data model. Second, the method lends itself to \emph{amortisation}: once a neural network is trained to emulate a process, it can be re-used repeatedly under very different data-model-scenarios. Third, even when generality and amortisation are not a concern, the method is still useful for scalability reasons. As shown in Section~\ref{sec:comp_efficiency}, guided diffusion will in many cases scale linearly with the dimension of the process. MCMC, on the other hand, often scales poorly with dimension.

There are some limitations of the proposed approach to spatial prediction. First, we found guided diffusion to approximate MCMC fairly well, with $\Pi$GDM achieving coverage closer to nominal than DPS on the cases we considered. However, both are slightly under-dispersed when compared to MCMC and both involve several approximations that, unlike MCMC, do not have any notion of asymptotic exactness. Second, vanilla guided diffusion freezes the process model, the process-model prior distributions, and the gridding of $\mathcal{D}$, and unlike MCMC, returns no posterior distribution over the process-model parameters $\thetab$. This may be unattractive in applications where the process-model parameters carry physical meaning. An appealing way forward is to train the ``prior'' diffusion model over both the process and the parameters \citep{Gloeckler_2024} or to combine guided diffusion with neural parameter estimation in an expectation-maximisation style framework \citep{Zammit_2025, Sainsbury-Dale_2025}. Finally, in the non-Gaussian case, the framework relies on having sufficient realisations from the process model in order to construct a network to estimate the score function required for reverse diffusion. For some applications where simulation is expensive (e.g., full Navier--Stokes type equations), this may be prohibitive. An alternative in this setting is to avoid score estimation altogether: \citet{Sharrock2026LatentFlow} write the process as a deterministic map of a latent innovation with a tractable reference distribution, so that conditioning requires solving a single reverse-time SDE and no training at all. This is attractive when simulations are scarce, although it requires the map from innovation to process to be available and differentiable.

\section*{Acknowledgements}

The authors would like to thank Alistair Manning for supplying the NAME simulations used in Section~\ref{sec:methane}. 

\section*{Declaration of generative AI and AI-assisted technologies in the manuscript preparation process}
During the preparation of this work the authors used Claude in order to help produce code and documentation from preliminary scripts and prompts. After using this tool, the authors reviewed and edited the content as needed and take full responsibility for the content of the published article.

\bibliography{references}

\newpage

\appendix

\counterwithin*{equation}{section}
\renewcommand{\theequation}{\Alph{section}.\arabic{equation}}
\renewcommand{\theHsection}{appendix.\Alph{section}}
\renewcommand{\theHequation}{appendix.\Alph{section}.\arabic{equation}}
\makeatletter
\renewcommand{\p@section}{Appendix~}
\renewcommand{\p@subsection}{Appendix~}
\makeatother

\section{Derivations of \eqref{eq:marginal}, \eqref{eq:reverse-sde} and \eqref{eq:dsm}}
\label{app:derivations}

For completeness, here we give derivations of the conditional distribution \eqref{eq:marginal}, the reverse-time SDE \eqref{eq:reverse-sde}, and the denoising score-matching objective \eqref{eq:dsm}. We obtain these from \citet{Anderson1982, Vincent2011, Song2021SDE}, but re-write them here in the notation and context of Section~\ref{sec:primer}.

\subsection{Conditional distribution \eqref{eq:marginal} and the coefficients $\alpha_t$ and $\sigma_t^2$}\label{subsec:transition-kernel}

The forward diffusion \eqref{eq:forward-sde} is a linear-Gaussian SDE with time-dependent coefficients. Conditional on $\Y_0$ the solution to the SDE is therefore an affine functional of Brownian motion, and it therefore suffices to compute its mean and variance.

Consider the integrating factor $\alpha_t^{-1} = \exp\!\bigl(\tfrac12\int_0^t \beta_s\intd s\bigr)$, so that $\tfrac{\intd}{\intd t}\alpha_t^{-1} = \tfrac12\beta_t\,
\alpha_t^{-1}$. It\^o's product rule for a deterministic function $\alpha_t^{-1}$ and stochastic process $\Y_t$ is given by $\intd\!\left(\alpha_t^{-1}\Y_t\right) = (\intd\alpha_t^{-1})\Y_t + \alpha_t^{-1}(\intd\Y_t)$. Therefore,
\begin{equation}
\label{eq:app-intfactor}
\intd\!\left(\alpha_t^{-1}\Y_t\right)
= \tfrac12\beta_t\,\alpha_t^{-1}\Y_t\,\intd t
  + \alpha_t^{-1}\!\left(-\tfrac12\beta_t \Y_t\,\intd t + \sqrt{\beta_t}\,\intd\W_t\right)
= \alpha_t^{-1}\sqrt{\beta_t}\,\intd\W_t.
\end{equation}
Integrating \eqref{eq:app-intfactor} from $0$ to $t$ and
multiplying through by $\alpha_t$ gives the explicit solution
\begin{equation}
\label{eq:app-ou-solution}
\Y_t = \alpha_t \Y_0 + \alpha_t \int_0^t \alpha_s^{-1}\sqrt{\beta_s}\,\intd\W_s .
\end{equation}
Since the It\^o integral has zero mean, taking conditional expectations in \eqref{eq:app-ou-solution} yields
\begin{equation}
\label{eq:app-mean}
\E[\Y_t \mid \Y_0] = \alpha_t \Y_0,
\end{equation}
which is the mean in \eqref{eq:marginal}. For the variance, note that the It\^o integral in
\eqref{eq:app-ou-solution} has independent components. We can therefore apply the It\^o isometry to each of these components to see that $\mathrm{Cov}(\Y_t \mid \Y_0) = \sigma_t^2 \I_n$ with
\begin{equation}
\label{eq:app-var-integral}
\sigma_t^2
= \alpha_t^2 \int_0^t \alpha_s^{-2}\,\beta_s\,\intd s .
\end{equation}
Since $\alpha_s^{-2} = \exp\!\bigl(\int_0^s \beta_u\,\intd u\bigr)$ satisfies
$\tfrac{\intd}{\intd s}\alpha_s^{-2} = \beta_s\,\alpha_s^{-2}$, we have that
\begin{equation}
\label{eq:app-var-eval}
\int_0^t \alpha_s^{-2}\,\beta_s\,\intd s
= \int_0^t \frac{\intd}{\intd s}\alpha_s^{-2}\,\intd s
= \alpha_t^{-2} - \alpha_0^{-2}
= \alpha_t^{-2} - 1 ,
\end{equation}
using $\alpha_0 = 1$. Substituting \eqref{eq:app-var-eval} into
\eqref{eq:app-var-integral} gives
\begin{equation*}
\sigma_t^2 = \alpha_t^2\bigl(\alpha_t^{-2} - 1\bigr) = 1 - \alpha_t^2 ,
\end{equation*}
which is the variance in \eqref{eq:marginal}.

\subsection{Reverse-time SDE \eqref{eq:reverse-sde}} \label{subsec:reverse-diffusion}

Write the forward diffusion \eqref{eq:forward-sde} in the generic It\^o form,
\begin{equation}
\label{eq:app-forward-generic}
\intd \Y_t = f(\Y_t, t)\, \intd t + g(t)\, \intd \W_t,
\end{equation}
where $f(\Y_t, t) \equiv -\tfrac{1}{2}\beta_t \Y_t,$ and $g(t) \equiv \sqrt{\beta_t}$. Let $p(\cdot)$ denote the marginal density of $\Y_t$. By the Fokker--Planck (forward) equation, $p(\cdot)$ evolves according to
\begin{equation}
\label{eq:app-fpe}
\partial_t p(\y_t)
= -\nabla \cdot \!\bigl[f(\y_t, t)\, p(\y_t)\bigr]
  + \tfrac{1}{2} g(t)^2\, \nabla^2 p(\y_t),
\end{equation}
where $\nabla \cdot$ and $\nabla^2$ denote the divergence and Laplacian in $\y_t$, respectively (subscripts are omitted for clarity). We seek a
diffusion process, run in reverse time, whose marginals coincide with $p(\cdot)$. Consider a
reverse-time SDE with the same diffusion coefficient $g(\cdot)$ and some drift $\bar f(\cdot)$,
\begin{equation}
\label{eq:app-reverse-generic}
\intd \Y_t = \bar f(\Y_t, t)\, \intd t + g(t)\, \intd \bar\W_t,
\end{equation}
where $\bar\W_t$ is a Brownian motion in reverse time and $\intd t$ is a negative
increment. Its marginals $q(\cdot)$ satisfy the Fokker--Planck (reverse) equation \citep{Anderson1982},
\begin{equation}
\label{eq:app-reverse-fpe}
\partial_t q(\y_t)
= -\nabla \cdot \!\bigl[\bar f(\y_t, t)\, q(\y_t)\bigr]
  - \tfrac{1}{2} g(t)^2\, \nabla^2 q(\y_t).
\end{equation}
For the reverse process to reproduce the required marginals we require $q(\cdot) \equiv p(\cdot)$ for all $t$. Rewriting the diffusion term of \eqref{eq:app-fpe} using $\nabla^2 p(\cdot) = \nabla \cdot (\nabla p(\cdot)) = \nabla \cdot \bigl(p(\cdot)\, \nabla \log p(\cdot)\bigr)$,
the forward equation \eqref{eq:app-fpe} becomes
\begin{equation}
\label{eq:app-fpe-rewritten}
\partial_t p(\y_t)
= -\nabla \cdot \!\Bigl[\bigl(f(\y_t,t) - g(t)^2\, \nabla \log p(\y_t)\bigr) p(\y_t)\Bigr]
  - \tfrac{1}{2} g^2(t)\, \nabla^2 p(\y_t) .
\end{equation}
Comparing \eqref{eq:app-fpe-rewritten} with \eqref{eq:app-reverse-fpe} and setting
$q(\cdot) = p(\cdot)$ identifies the reverse drift,
\begin{equation}
\label{eq:app-reverse-drift}
\bar f(\y_t, t) = f(\y_t, t) - g(t)^2\, \nabla \log p(\y_t).
\end{equation}
Substituting $f(\y_t, t) = -\tfrac{1}{2}\beta_t \y_t$ and $g(t)^2 = \beta_t$ into
\eqref{eq:app-reverse-drift} and \eqref{eq:app-reverse-generic} gives \eqref{eq:reverse-sde}. 

\subsection{Denoising score-matching objective \eqref{eq:dsm}} \label{subsec:Score}

Here we give the derivation for the loss \eqref{eq:dsm} used to train a neural network to approximate the score $\nabla \log p(\Y_t)$; the derivation for the loss \eqref{eq:dsm2} is similar in many respects, and not shown for brevity.

The starting point of the derivation is the explicit score-matching loss
\begin{equation}
\label{eq:app-esm}
J_{\mathrm{ESM}}(\phib)
= \E_{\Y_t}
  \left[\bigl\| \sphi(\Y_t, t) - \nabla \log p(\Y_t) \bigr\|^2\right],
\end{equation}
which is minimised by the true score $\sphi(\Y_t, t) = \nabla \log p(\Y_t)$ for all $t$ (as in \ref{subsec:reverse-diffusion}, here we omit subscripts on the gradient operator for clarity). Equation \eqref{eq:app-esm} however cannot be evaluated directly because $\nabla \log p(\Y_t)$ is not known in general. Following \citet{Vincent2011}, we show that it is equivalent, up to a constant independent of $\phib$, to the denoising score-matching loss
\begin{equation}
\label{eq:app-dsm}
J_{\mathrm{DSM}}(\phib)
= \E_{\Y_0}\E_{\Y_t \mid \Y_0}
  \left[\bigl\| \sphi(\Y_t, t) - \nabla \log p(\Y_t \mid \Y_0) \bigr\|^2\right],
\end{equation}
for all $t$, in which the conditional score $\nabla \log p(\Y_t \mid \Y_0)$ is available in
closed form. Expanding the squared norms in \eqref{eq:app-esm} and \eqref{eq:app-dsm}, the
$\|\sphi(\Y_t, t)\|^2$ terms are identical and the terms not involving $\sphi(\Y_t, t)$ are constant in
$\phib$; the two objectives therefore differ only through their cross terms. Since $p(\y_t)\, \nabla \log p(\y_t) = \nabla p(\y_t)$, the cross term
of \eqref{eq:app-esm} is
\begin{equation}
\label{eq:app-cross}
\E_{\Y_t}\!\bigl[\sphi(\Y_t, t)^\top \nabla \log p(\Y_t)\bigr]
= \int \sphi(\y_t, t)^\top \nabla p(\y_t)\, \intd\y_t.
\end{equation}
Now, since $p(\y_t) = \int p(\y_0)\, p(\y_t \mid \y_0)\, \intd\y_0$, we have that
\begin{equation}
\label{eq:app-marg-grad}
\nabla p(\y_t)
= \int p(\y_0)\, \nabla p(\y_t \mid \y_0)\, \intd\y_0
= \int p(\y_0)\, p(\y_t \mid \y_0)\, \nabla \log p(\y_t \mid \y_0)\, \intd\y_0 .
\end{equation}
Substituting \eqref{eq:app-marg-grad} into \eqref{eq:app-cross} yields
\begin{equation*}
\E_{\Y_t}\!\bigl[\sphi(\Y_t, t)^\top \nabla \log p(\Y_t)\bigr]
= \E_{\Y_0}\E_{\Y_t \mid \Y_0}
  \!\bigl[\sphi(\Y_t, t)^\top \nabla \log p(\Y_t \mid \Y_0)\bigr].
\end{equation*}
Hence
$J_{\mathrm{ESM}}(\phib) = J_{\mathrm{DSM}}(\phib) + \mathrm{const}$, and the two losses
share the same minimiser for each $t$ \citep{Vincent2011}.

To find an optimal $\phib$ for all $t$, it is sufficient to minimise $J_{\mathrm{DSM}}(\phib)$ uniformly over $t$. Integrating \eqref{eq:app-dsm} with respect to time gives the objective $\mathcal{L}_\text{S}(\phib)$ in \eqref{eq:dsm}.

\section{$\Pi$GDM approximations for non-linear data models}
\label{app:pigdm-nonlinear}

This appendix details the approximations used to evaluate the noised likelihood \eqref{eq:pigdm-noised} for $\Pi$GDM with the count, right-censoring and area-exceedance data models of Section~\ref{subsec:nonlinear}.

\subsection{Poisson data}
\label{app:pigdm-poisson}

Under the Poisson data model \eqref{eq:poisson-obs} the noised likelihood \eqref{eq:pigdm-noised} is not available in closed form. To make the integral tractable, since $\mathbf{H}$ is an incidence matrix we approximate each factor $p(Z_i \mid Y_{0,i})$ by a Gaussian in $Y_{0,i}$. A second-order Taylor expansion of $\log p(Z_i \mid Y_{0,i})$ about the Tweedie estimate $\hat{Y}_{0,i}(\Y_t)$, with $u_i \equiv Y_{0,i} - \hat{Y}_{0,i}(\Y_t)$, gives
\[
\log p(Z_i \mid Y_{0,i}) \approx \log p(Z_i \mid \hat{Y}_{0,i}(\Y_t)) + g_i\,u_i - \tfrac{1}{2}\Lambda_{ii}\,u_i^2,
\]
where 
\[
g_i \equiv \left.\frac{\partial\, \log p(Z_i \mid Y_{0,i})}{\partial Y_{0,i}}\right|_{Y_{0,i} = \hat{Y}_{0,i}(\Y_t)}, \qquad
\Lambda_{ii} \equiv -\left.\frac{\partial^2 \log p(Z_i \mid Y_{0,i})}{\partial Y_{0,i}^2}\right|_{Y_{0,i} = \hat{Y}_{0,i}(\Y_t)},
\]
are the log-likelihood first and second derivatives with respect to $Y_{0,i}$, respectively. Completing the square in $u_i$,
\[
g_i\,u_i - \tfrac{1}{2}\Lambda_{ii}\,u_i^2 = -\tfrac{\Lambda_{ii}}{2}\bigl(u_i - \Lambda_{ii}^{-1} g_i\bigr)^2 + \tfrac{g_i^2}{2\Lambda_{ii}},
\]
whose final term is constant in $Y_{0,i}$. Introducing the pseudo-response
\[
\tilde{Z}_i \equiv \hat{Y}_{0,i}(\Y_t) + \Lambda_{ii}^{-1} g_i,
\]
gives $p(Z_i \mid Y_{0,i}) \propto \exp\bigl(-\tfrac{\Lambda_{ii}}{2}(Y_{0,i} - \tilde{Z}_i)^2\bigr)$. This is, up to a factor independent of $Y_{0,i}$, the density of a Gaussian pseudo-response
\[
\tilde{Z}_i \mid Y_{0,i} \sim \Gau\,\bigl(Y_{0,i},\, \Lambda_{ii}^{-1}\bigr).
\]

Stacking the $m$ observed cells, the pseudo-responses satisfy $\tilde{\Z} = \Hmat\Y_0 + \boldsymbol{\eta}$ with ${\boldsymbol{\eta} \sim \Gau(\zerov, \boldsymbol{\Lambda}^{-1})}$ and $\boldsymbol{\Lambda}$ the diagonal matrix comprising the $\Lambda_{ii}, i = 1,\dots,m$. The noised likelihood therefore follows from Section~\ref{subsec:linear}:
\begin{equation}
\label{eq:pigdm-poisson-noised}
\tilde{\Z} \mid \Y_t \;\approx\; \Gau\bigl(\Hmat\Yhat(\Y_t),\, \Hmat\mathbf{V}_t\Hmat^\top + \boldsymbol{\Lambda}^{-1}\bigr),
\end{equation}
and, as in the linear-Gaussian case, the guidance gradient follows by differentiating the resulting log-density through $\Yhat(\Y_t)$.

\subsection{Right censoring}
\label{app:pigdm-censoring}

Recall that under the right-censoring data model, 
\begin{equation}
\label{eq:censor-tail2}
\log p(\Z, \mathbf{M} \mid \Y_0)
= \sum_{i: M_i = 0} \left[-\frac{1}{2}\left(\frac{Z_i - Y_{0,i}}{\sigma_\epsilon}\right)^{\!2}\right]
+ \sum_{i: M_i = 1} \log \Phi\!\left(\frac{Y_{0,i} - c}{\sigma_\epsilon}\right),
\end{equation}
up to an additive constant independent of $\Y_0$. Since the data are pixel-referenced and are conditionally independent given the latent process, we can write ${p(\Z, \mathbf{M} \mid \Y_0) = \prod_{i=1}^n p_i(Z_i, M_i  \mid Y_{0,i})}$ where the factors are defined from \eqref{eq:censor-tail2}. Now, assume that we can write $\mathbf{V}_t = v_t \I_n$ (see the discussion below \eqref{eq:pigdm-surrogate}); then the integral in \eqref{eq:pigdm-noised} is a product of one-dimensional integrals,
\[
p(\Z, \mathbf{M} \mid \Y_t) \;\approx\; \prod_{i=1}^n \int p_i(Z_i, M_i \mid Y_{0,i})\, \Gau\bigl(Y_{0,i};\, \hat{Y}_{0,i}(\Y_t),\, v_t\bigr) \intd Y_{0,i}.
\]
For an uncensored cell ($M_i = 0$), the noised likelihood component is
\[
\int \Gau\bigl(Z_i; Y_{0,i}, \sigma_\epsilon^2\bigr)\, \Gau\bigl(Y_{0,i}; \hat{Y}_{0,i}(\Y_t), v_t\bigr) \intd Y_{0,i}
= \Gau\bigl(Z_i;\, \hat{Y}_{0,i}(\Y_t),\, \sigma_\epsilon^2 + v_t\bigr),\qquad i: M_i = 0.
\]
For a censored cell ($M_i = 1$), note that $\Phi\!\left((Y_{0,i} - c)/\sigma_\epsilon\right) = \Pr(\epsilon_i \le Y_{0,i} - c \mid Y_{0,i})$. Averaging over $Y_{0,i} \sim \Gau(\hat{Y}_{0,i}(\Y_t), v_t)$ and applying the law of total probability,
\[
\int \Phi\!\left(\frac{Y_{0,i} - c}{\sigma_\epsilon}\right) \Gau\bigl(Y_{0,i}; \hat{Y}_{0,i}(\Y_t), v_t\bigr) \intd Y_{0,i}
= \Pr(\epsilon_i \le Y_{0,i} - c)
= \Pr(Y_{0,i} - \epsilon_i \ge c).
\]
As $Y_{0,i}$ and $\epsilon_i$ are independent and Gaussian, $Y_{0,i} - \epsilon_i \sim \Gau\bigl(\hat{Y}_{0,i}(\Y_t),\, \sigma_\epsilon^2 + v_t\bigr)$, and hence
\[
\Pr(Y_{0,i} - \epsilon_i \ge c) = \Phi\!\left(\frac{\hat{Y}_{0,i}(\Y_t) - c}{\sqrt{\sigma_\epsilon^2 + v_t}}\right).
\]
This leads to the noised likelihood component
\[
\int \Phi\!\left(\frac{Y_{0,i} - c}{\sigma_\epsilon}\right) \Gau\bigl(Y_{0,i}; \hat{Y}_{0,i}(\Y_t), v_t\bigr) \intd Y_{0,i}
= \Phi\!\left(\frac{\hat{Y}_{0,i}(\Y_t) - c}{\sqrt{\sigma_\epsilon^2 + v_t}}\right),\qquad i: M_i = 1.
\]

Taking logs of both components and evaluating at $\Yhat(\Y_t)$ we obtain the noised log-likelihood
\begin{equation}
\label{eq:pigdm-censor-noised}
\log p(\Z, \mathbf{M} \mid \Y_t)
\;\approx\;
\sum_{i: M_i = 0} \left[-\frac{1}{2}\left(\frac{Z_i - \hat{Y}_{0,i}(\Y_t)}{\sqrt{\sigma_\epsilon^2 + v_t}}\right)^{\!2}\right]
+ \sum_{i: M_i = 1} \log \Phi\!\left(\frac{\hat{Y}_{0,i}(\Y_t) - c}{\sqrt{\sigma_\epsilon^2 + v_t}}\right),
\end{equation}
up to an additive constant.  Comparing \eqref{eq:pigdm-censor-noised} to \eqref{eq:censor-tail2} we see that conditioning on $\Y_t$ instead of $\Y_0$ effectively inflates the variance, as expected.

\subsection{Area-exceedance functional}
\label{app:pigdm-area}

To incorporate the area-exceedance functional with $\Pi$GDM, we first construct a Gaussian approximation of the data model. Define the normalised discrete weights 
$$w_k(\Y_0) = \frac{\exp\bigl(-(k - h(\Y_0))^2/(2\xi^2)\bigr)}{\sum_{j=0}^n \exp\bigl(-(j - h(\Y_0))^2/(2\xi^2)\bigr)} \quad  k \in \{0,\ldots,n\},$$
and set
\[
\mu(\Y_0) \equiv \E[Z \mid \Y_0] = \sum_{k=0}^n k\, w_k(\Y_0),
\qquad
\gamma(\Y_0) \equiv \mathrm{Var}(Z \mid \Y_0) = \sum_{k=0}^n k^2 w_k(\Y_0) - \mu(\Y_0)^2.
\]
Then $p(Z \mid \Y_0) \approx \Gau\bigl(\mu(\Y_0), \gamma(\Y_0)\bigr)$. Both $\mu(\cdot)$ and $\gamma(\cdot)$ depend on $\Y_0$ through the ``soft count'' $h(\Y_0) = \sum_{i=1}^n \sigma_\kappa(Y_{0,i})$ where, recall, $\sigma_\kappa(Y_{0,i}) \equiv \text{expit}(\kappa(Y_{0,i} - c))$. Because $\xi^2$ is small, this discrete Gaussian concentrates almost all of its mass within one unit of $h(\Y_0)$. We therefore approximate $\gamma(\Y_0) \approx \gamma(\Yhat(\Y_t))$, which is correspondingly a small variance controlled by $\xi^2$. Under this approximation for the data model, \eqref{eq:pigdm-noised} becomes
\[
p(Z \mid \Y_t) \;\approx\; \int \Gau\bigl(Z; \mu(\Y_0), \gamma(\Yhat(\Y_t))\bigr)\, \Gau\bigl(\Y_0; \Yhat(\Y_t), \mathbf{V}_t\bigr) \intd \Y_0,
\]
which is intractable since $\mu(\Y_0)$ depends non-linearly on $\Y_0$.  We recover tractability by linearising the $\mu(\cdot)$ about $\Yhat(\Y_t)$. Specifically, let $$J_i(\Y_0) \equiv \partial \mu(\Y_0)/\partial Y_{0,i} = \bigl(\intd \mu(\Y_0)/\intd h(\Y_0)\bigr)\,\partial h(\Y_0)/\partial Y_{0,i}.$$ Differentiating the weights $w_k$ with respect to $h$ gives $\partial w_k/\partial h = w_k\,(k - \mu)/\xi^2$, so that
\[
\frac{\intd \mu}{\intd h} = \sum_{k=0}^n k\,\frac{\partial w_k}{\partial h}
 = \frac{1}{\xi^2}\Bigl(\sum_{k=0}^n k^2 w_k - \mu(\Y_0)^2\Bigr) = \frac{\gamma(\Y_0)}{\xi^2},
\]
Combining this with $\partial h(\Y_0)/\partial Y_{0,i} = \kappa\,\sigma_\kappa(Y_{0,i})\bigl(1 - \sigma_\kappa(Y_{0,i})\bigr)$ yields
\begin{equation}
\label{eq:pigdm-area-jacobian}
J_i(\Y_0) = \frac{\gamma(\Y_0)}{\xi^2}\,\kappa\, \sigma_\kappa(Y_{0,i})\bigl(1 - \sigma_\kappa(Y_{0,i})\bigr).
\end{equation}
A first-order Taylor expansion of the mean is then given by,
\[
\mu(\Y_0) \;\approx\; \mu\bigl(\Yhat(\Y_t)\bigr) + \mathbf{J}^\top\bigl(\Yhat(\Y_t)\bigr)\bigl(\Y_0 - \Yhat(\Y_t)\bigr),
\]
where $\mathbf{J}(\cdot) \equiv \bigl(J_1(\cdot), \ldots, J_n(\cdot)\bigr)^\top$. We can now apply the ``re-parameterisation trick'' to write
$$
Z = \mu(\Yhat(\Y_t)) + \mathbf{J}(\Yhat(\Y_t))^\top\,(\Y_0 - \Yhat(\Y_t)) + \eta, \qquad \eta \sim \Gau(0, \gamma(\Yhat(\Y_t))),  
$$
so that, when substituting in for $\Y_0 \sim \Gau(\Yhat(\Y_t), \mathbf{V}_t)$, we have that
\begin{equation}
\label{eq:pigdm-area-noised}
Z \mid \Y_t \;\approx\; \Gau\,\!\Bigl(\mu\bigl(\Yhat(\Y_t)\bigr),\; \tilde{v}_t\Bigr),
\qquad
\tilde{v}_t \equiv \mathbf{J}(\Yhat)^\top\, \mathbf{V}_t\, \mathbf{J}(\Yhat) + \gamma\bigl(\Yhat(\Y_t)\bigr).
\end{equation}

\clearpage

\counterwithout*{equation}{section}
\makeatletter
\let\p@section\@empty
\let\p@subsection\@empty
\makeatother
\setcounter{section}{0}
\setcounter{equation}{0}
\setcounter{figure}{0}
\setcounter{table}{0}
\renewcommand{\thesection}{S.\arabic{section}}
\renewcommand{\theequation}{S.\arabic{equation}}
\renewcommand{\thefigure}{S.\arabic{figure}}
\renewcommand{\thetable}{S.\arabic{table}}
\renewcommand{\theHsection}{supp.\arabic{section}}
\renewcommand{\theHequation}{supp.\arabic{equation}}
\renewcommand{\theHfigure}{supp.\arabic{figure}}
\renewcommand{\theHtable}{supp.\arabic{table}}

\begin{center}
  {\Large\bfseries Supplementary Material}
\end{center}

\section{Computational details}
\label{app:compute}
For the forward process \eqref{eq:forward-sde} we use a variance-preserving schedule with linear $\beta_t = \beta_0 + {(\beta_1-\beta_0)t}$, $\beta_0 = 0.1$, $\beta_1 = 20$, and  $T = 1$, so that $\alpha_T \approx 0.007$ and hence $\Y_T  \mathrel{\dot\sim} \Gau(\zerov,\I_n)$.

To numerically integrate when reverse sampling, we discretise \eqref{eq:reverse-sde} on a grid $T=t_T>t_{T-1}>\cdots>t_0=0$ and, on this grid, we use an update of the form
\begin{equation}
\label{eq:euler-step}
\Y_{t_{k-1}} = \mu_{\phib}(\Y_{t_k},t_k) + \tilde{\sigma}_{t_k}\boldsymbol{\omega}_{t_k}, \qquad \boldsymbol{\omega}_{t_k}\sim\Gau(\zerov,\I_n),
\end{equation}
for $k = T,\dots,1$,  where $\mu_{\phib}(\Y_{t_k},t_k)$ corresponds to the reverse-time dynamics in \eqref{eq:reverse-sde}, evaluated using $\rphi(\Y_{t_k},t_k)$, and $\tilde{\sigma}_{t_k}$ is a step-dependent scaling specified by the diffusion schedule. We construct \eqref{eq:euler-step} using the Euler--Maruyama scheme of \citet{Song2021SDE} as implemented by \citet{moss2026conditioning}, on a time grid $\{t_k\}$ spaced uniformly in the log signal-to-noise ratio $\log(\alpha_t^2/\sigma_t^2)$. 

At every reverse step we soft-norm-clip the guidance contribution to the reverse drift via a $\tanh$ squashing saturating at $100$. For the first Gaussian process example in Section~\ref{subsec:gp-validation} the norm stayed below the bound for both DPS and $\Pi$GDM. For DPS in the other experiments the raw guidance norm grew by several orders of magnitude near $t=0$, and clipping was necessary. Guidance with $\Pi$GDM is better conditioned and rarely required clipping.  

In Section~\ref{subsec:gp-validation} the Gaussian-process score is available in closed form, so no neural network is needed. For the non-Gaussian process models, as score network we used a U-Net with circular padding for the periodic Ginzburg–Landau domain (Section~\ref{subsec:gl-prior}) and zero padding for the bounded methane domain (Section~\ref{sec:methane}). The U-Nets were trained using \eqref{eq:dsm2} with Adam (learning rate $2\times10^{-4}$, batch size $128$, exponential moving average with decay $0.999$). For more details on the neural networks, please see the supplementary code.

A convolutional score network requires a full rectangular array, whereas the methane flux field of Section~\ref{sec:methane} is defined only on the $n = 122$ land cells. We therefore embed each simulated flux field in the enclosing $20 \times 20$ longitude--latitude array, and fill the remaining $278$ (sea and out-of-domain) cells with the median flux of the simulated ensemble; the higher-resolution configurations of Section~\ref{sec:comp_efficiency} are treated identically. This choice does not affect the inferences: the transport operator $\Hmat$ in \eqref{eq:methane-obs} has zero columns at the inactive cells, and every predictive sample is restricted to the land cells before any diagnostic is computed.

\section{Comparison with neural conditional simulation}
\label{app:ncs}

The guided-diffusion framework defers the choice of data model to prediction time, whereas NCS \citep{Walchessen2024NCS} requires the data model to be fixed before its conditional score network $\tilde{\mathbf{r}}_{\phib}(\Y_t, \Z, t)$ is trained (Section~\ref{sec:primer}). To illustrate the practical consequence of this difference, we conduct an experiment on the Ginzburg--Landau amplitude field of Section~\ref{subsec:gl-prior}, using the same held-out realisations and the same training fields as the guided approaches, so that the two approaches differ only in how the conditioning information is incorporated: NCS learns a single conditional score network, whereas guided diffusion pairs the unconditional process-model score with the data-model likelihood at prediction time.

We train an NCS network on a missing-at-random (MAR) Gaussian data model. For each simulated field, a random proportion of the grid cells are retained (where the observed proportion is uniformly generated between 0.02 and 0.10). Gaussian measurement error of known standard deviation $\sigma_\epsilon = 0.1$ is then added, and the network is trained on the corresponding conditional score-matching objective \eqref{eq:ncs-loss}. We then apply the trained network in two settings. In the first setting the data are generated with missingness at random, with an observed proportion of $5\%$. In the second setting, the data instead arise from the upper-tail censoring model of Section~\ref{subsec:gl-prior}, in which a cell is observed only when its noisy measurement $Z_i < c$, $c = 1$. In both settings we compare NCS against guided diffusion (DPS and $\Pi$GDM) with the appropriate data-model, and we report RMSE, $90\%$ coverage, ES, and CRPS over $50$ realisations. For the censoring setting we report these diagnostics over the censored cells.

\begin{figure}[h!]
\centering
\includegraphics[width=\textwidth]{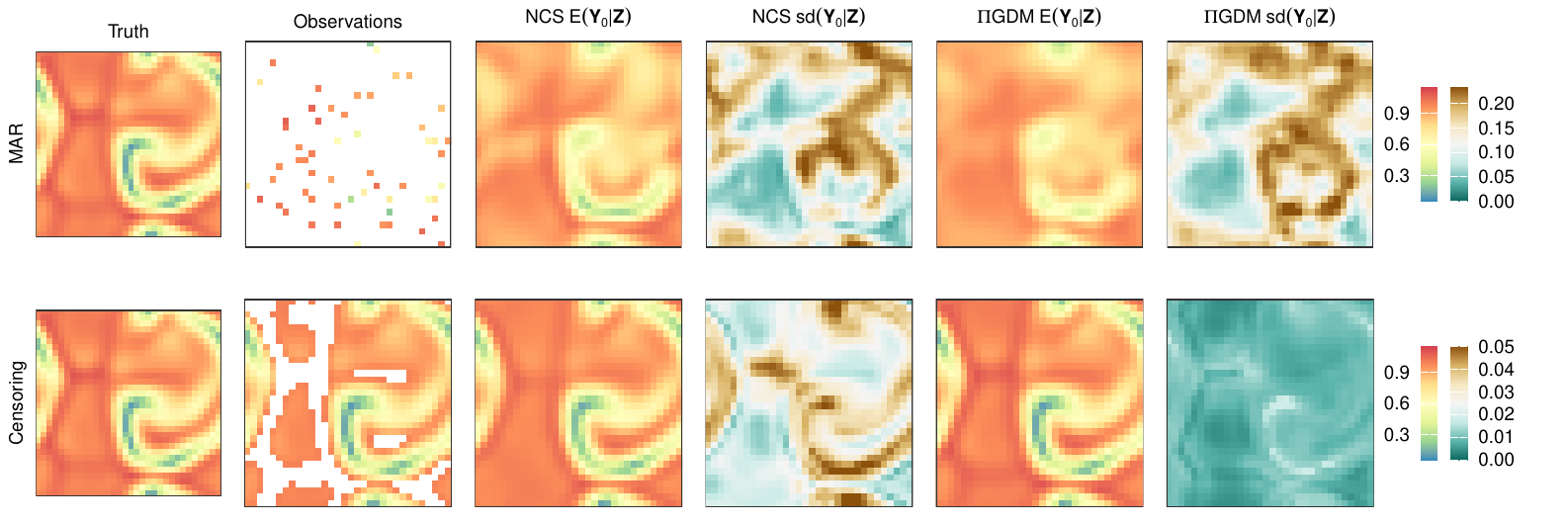}
\caption{Predictions of a Ginzburg--Landau amplitude field for one realisation under the MAR model (top) and the upper-tail censoring model (bottom). Columns show the true field, the observations (missing or censored cells in white), and the predictive mean and standard deviation for NCS and for $\Pi$GDM. The quantitative differences between the methods over several realisations are reported in Table~\ref{tab:ncs}.}
\label{fig:ncs}
\end{figure}

Table~\ref{tab:ncs} summarises the results, and Figure~\ref{fig:ncs} shows predictions for one realisation. Under the matched MAR model, NCS is similar to guided diffusion. However, under the censoring model, the predictive accuracy of NCS deteriorates, as expected. The reason for this deterioration is that the NCS network is trained for data that are missing at random with a missingness proportion of between 0.90 and 0.98, and that have a measurement error standard deviation of $\sigma_\epsilon = 0.1$. The guided samplers, on the other hand, exploit the correct censoring likelihood \eqref{eq:censor-tail} without the need for any retraining. Therefore, NCS performs well and arguably should be used when the data model is known before training time, but cannot accommodate a change in the data model without retraining. This is a limitation that the guided-diffusion framework removes.

\begin{table}[h!]
\centering
\caption{Comparison of neural conditional simulation (NCS) with guided diffusion (DPS and $\Pi$GDM) on the Ginzburg--Landau amplitude field. The NCS network is trained on a missing-at-random (MAR) Gaussian data model and applied, without retraining, both to MAR data and to the upper-tail censoring model of Section~\ref{subsec:gl-prior}. RMSE (of the predictive mean), $90\%$-interval coverage, ES, and CRPS are computed against the true field, as averages over $50$ realisations; for the censoring setting the diagnostics are computed over the censored cells.}
\label{tab:ncs}
\vspace{1em}
\begin{tabular}{llrrrr}
\hline
Scenario & Method & RMSE & Cov.\ (0.90) & ES & CRPS \\
\hline
MAR & NCS & 0.154 & 0.893 & 3.47 & 0.079 \\
 & DPS & 0.156 & 0.856 & 3.54 & 0.081 \\
 & $\Pi$GDM & 0.158 & 0.903 & 3.57 & 0.081 \\
\hline
Censoring & NCS & 0.040 & 0.739 & 0.39 & 0.022 \\
 & DPS & 0.010 & 0.816 & 0.10 & 0.006 \\
 & $\Pi$GDM & 0.011 & 0.860 & 0.11 & 0.006 \\
\hline
\end{tabular}

\end{table}

\clearpage

\section{Additional Figures}
\label{app:additional-figures}

\begin{figure}[h!]
\centering
\includegraphics[width=\textwidth]{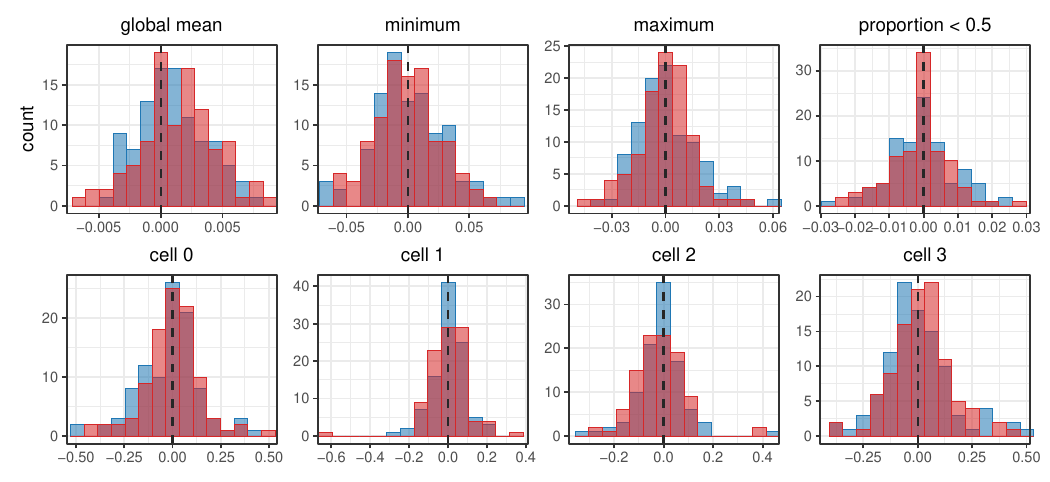}
\caption{Functional residuals for the areal aggregation experiment of Section~\ref{subsec:gl-prior} ($100$ independent realisations) for DPS (blue) and $\Pi$GDM (red). The dashed line is at zero.}
\label{fig:app-pred-functionals-aggregates}
\end{figure}

\begin{figure}[h!]
\centering
\includegraphics[width=\textwidth]{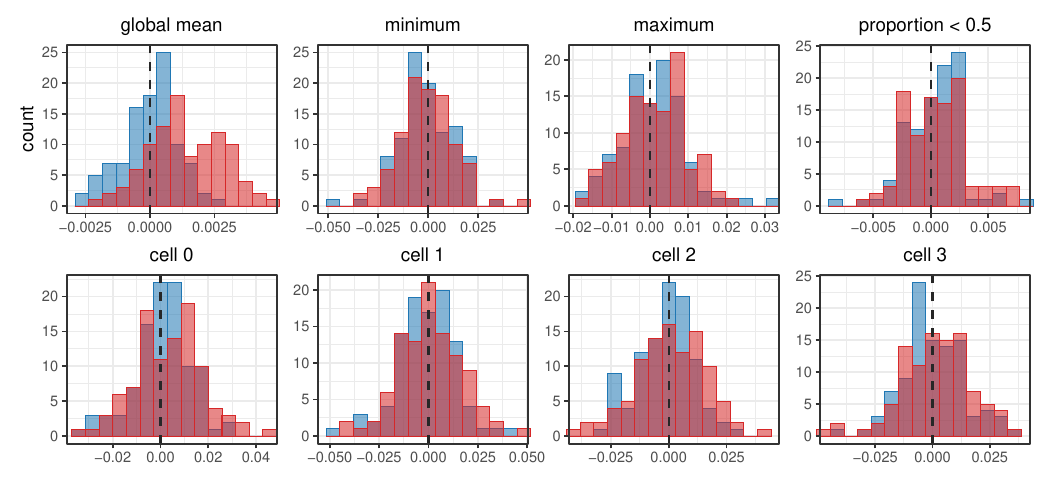}
\caption{Same as Figure~\ref{fig:app-pred-functionals-aggregates} but for the censored data experiment.}
\label{fig:app-pred-functionals-censored}
\end{figure}

\begin{figure}[h!]
\centering
\includegraphics[width=\textwidth]{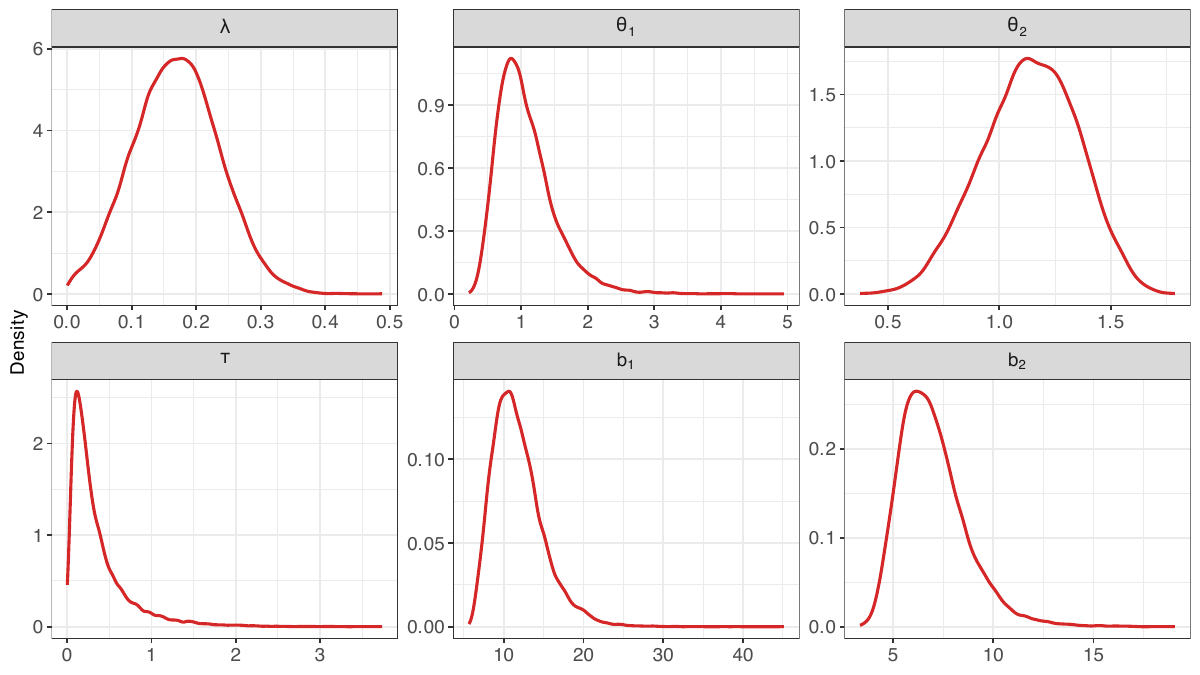}
\caption{Prior distributions used in the methane flux experiment \eqref{eq:methane-hier}, elicited from the inventory alone. Panels show the induced model-scale parameters $(\lambda, \theta_1, \theta_2, \tau, b_1, b_2)$ obtained from independent informative priors on the natural-scale parameters $(\lambda, \log\theta_1, \mathrm{logit}(\theta_2/2), \log\tilde{b}_1, \log\tilde{b}_2, \log c_{\mathrm{V}})$ via \eqref{eq:methane-reparam}.}
\label{fig:methane-prior-hyper}
\end{figure}

\begin{figure}[h!]
  \centering
  \includegraphics[width=\textwidth]{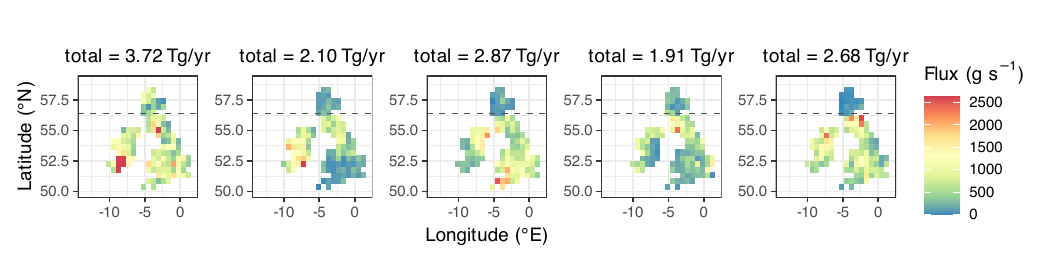}
  \vspace{-0.3in} 
  \caption{ Five unconditional draws from $p(\mathbf{Y}_0)$ via reverse diffusion, each annotated with the domain flux total. For context, the methane inventory flux is shown in Figure~\ref{fig:methane-means-and-se} and has a total of $2.73$~Tg\,yr$^{-1}$.}
  \label{fig:methane-prior}
  \end{figure}

  \begin{table}[h!]
    \centering
    \caption{Region-level total methane flux (g\,s$^{-1}$): true (inventory) value, and predictive mean $\pm$ predictive standard deviation for the two guided samplers (DPS and $\Pi$GDM) and MCMC (Section~\ref{subsec:methane-results}). \vspace{0.1in}}
    \label{tab:methane-countries}
    \begin{tabular}{lcccc}
\hline
Territory & Truth & DPS & $\Pi$GDM & MCMC \\
\hline
England & $45{,}604$ & $45{,}699 \pm 1{,}333$ & $45{,}589 \pm 1{,}491$ & $45{,}184 \pm 1{,}614$ \\
Scotland & $10{,}682$ & $13{,}391 \pm 1{,}561$ & $13{,}780 \pm 1{,}809$ & $13{,}248 \pm 1{,}713$ \\
Wales & $7{,}888$ & $8{,}246 \pm 606$ & $8{,}059 \pm 671$ & $8{,}102 \pm 739$ \\
Northern Ireland & $6{,}318$ & $5{,}484 \pm 861$ & $5{,}411 \pm 935$ & $5{,}578 \pm 956$ \\
Ireland & $16{,}097$ & $14{,}499 \pm 1{,}163$ & $14{,}529 \pm 1{,}356$ & $14{,}533 \pm 1{,}424$ \\
\hline
\end{tabular}

    \end{table}

\begin{figure}[p]
\centering
\includegraphics[width=\textwidth]{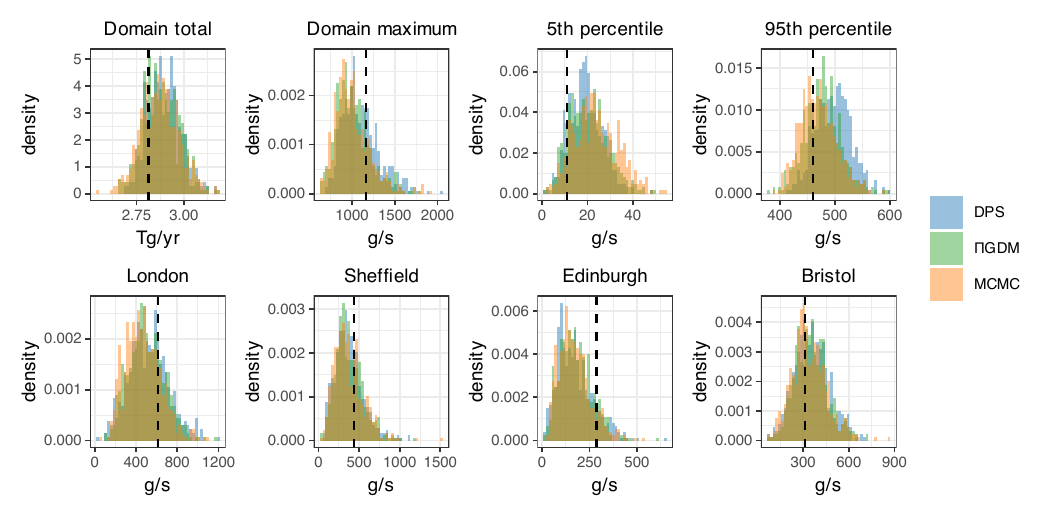}
\caption{As Figure~\ref{fig:methane-functionals} but for $n = 487$: predictive distributions of eight functionals of the flux field for DPS, $\Pi$GDM, and MCMC.}
\label{fig:hires2-functionals}
\end{figure}

\begin{figure}[p]
\centering
\includegraphics[width=\textwidth]{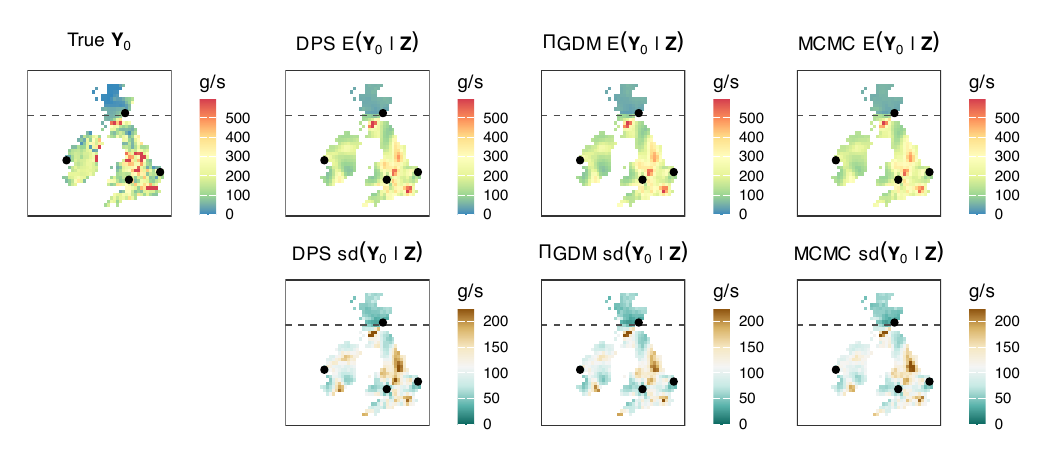}
\caption{As Figure~\ref{fig:methane-means-and-se} but for $n = 487$: per-cell predictive mean and predictive standard deviation for DPS, $\Pi$GDM, and MCMC.}
\label{fig:hires2-mean-sd}
\end{figure}

\begin{figure}[p]
\centering
\includegraphics[width=\textwidth]{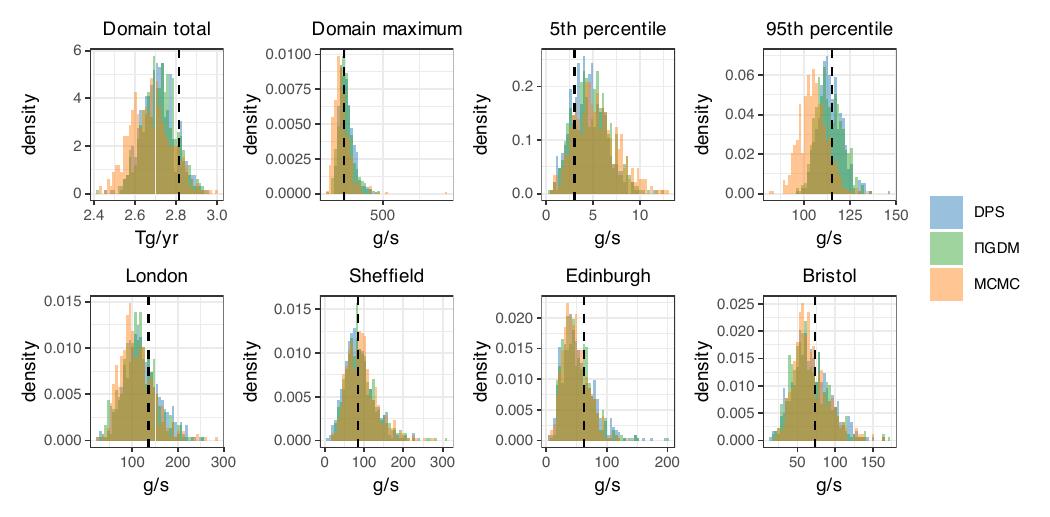}
\caption{As Figure~\ref{fig:methane-functionals} but for $n = 1948$: predictive distributions of eight functionals of the flux field for DPS, $\Pi$GDM, and MCMC.}
\label{fig:ultra-functionals}
\end{figure}

\begin{figure}[p]
\centering
\includegraphics[width=\textwidth]{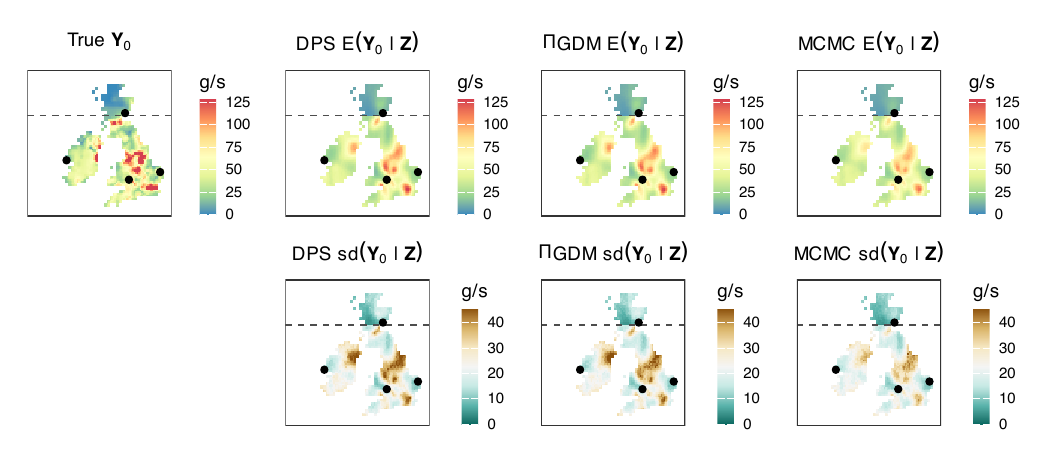}
\caption{As Figure~\ref{fig:methane-means-and-se} but for $n = 1948$: per-cell predictive mean and predictive standard deviation for DPS, $\Pi$GDM, and MCMC.}
\label{fig:ultra-mean-sd}
\end{figure}

\clearpage

\newpage

\end{document}